\documentclass[12pt,draftcls,onecolumn]{IEEEtran}
\usepackage{amsmath}
\usepackage{amsfonts}
\usepackage{amssymb}
\usepackage{amsthm}
\usepackage[utf8]{inputenc}
\DeclareUnicodeCharacter{00A0}{ }
\usepackage{graphicx}
\usepackage{graphics}
\usepackage{float}
\usepackage{tikz}
\usetikzlibrary{shapes,snakes,patterns}
\usepackage{epstopdf}
\renewcommand{\vec}[1]{\mathbf{#1}}
\usepackage[justification=centering]{caption}
\usepackage{enumerate} 
\usepackage{cite}
\usepackage{placeins}
\newcommand{\vast}{\bBigg@{4}}
\newcommand{\Vast}{\bBigg@{5}}
\newtheorem{theorem}{Theorem}[]

\usepackage{suffix}
\usepackage{mathtools}
\usepackage{enumitem}
\usepackage{xfrac}
\usepackage{relsize}

\theoremstyle{definition}

\usepackage[list=true]{subcaption}
\DeclarePairedDelimiterX\MeijerM[3]{\lparen}{\rparen}%
{\begin{smallmatrix}#1 \\ #2\end{smallmatrix}\delimsize\vert\,#3}

\newcommand\MeijerG[8][]{%
  G^{\,#2,#3}_{#4,#5}\MeijerM[#1]{#6}{#7}{#8}}

\WithSuffix\newcommand\MeijerG*[7]{%
  G^{\,#1,#2}_{#3,#4}\MeijerM*{#5}{#6}{#7}}

\begin{document}

\title{Approximate Random Matrix Models for Generalized Fading MIMO Channels}

\author{Muralikrishnan Srinivasan  \hspace{0.8cm} Sheetal Kalyani   \\
 \hspace{-0.5 cm}Department of Electrical Engineering,\\
  \hspace{-0.8 cm} Indian Institute of Technology, Madras, \\
 \hspace{-1cm} Chennai, India 600036.\\
 \hspace{-1cm} \{ee14d206,skalyani\}@ee.iitm.ac.in\\
 }
 \maketitle
 
 \begin{abstract}
Approximate random matrix models for $\kappa-\mu$ and $\eta-\mu$ faded multiple input multiple output (MIMO) communication channels are derived in terms of a complex Wishart matrix. The proposed approximation has the least Kullback-Leibler (KL) divergence from the original matrix distribution. The utility of the results are demonstrated in a) computing the average capacity/rate expressions of $\kappa-\mu$/$\eta-\mu$ MIMO systems b) computing outage probability (OP) expressions for maximum ratio combining (MRC) for $\kappa-\mu$/$\eta-\mu$ faded MIMO channels c) ergodic rate expressions for zero-forcing (ZF) receiver in an uplink single cell massive MIMO scenario with low resolution analog-to-digital converters (ADCs) in the antennas. These approximate expressions are compared with Monte-Carlo simulations and a close match is observed. 
\end{abstract}


\section{Introduction}
\IEEEPARstart{T}{he} need for high data rates has been one of the driving factors for the evolution of the wireless systems from Single Input Single Output (SISO) systems to Multiple Input Multiple Output (MIMO) systems. MIMO systems are being used increasingly in modern wireless standards and it is imperative to study the channel capacity and other quality of service (QoS) metrics of such systems. But metrics likes capacity depend not only on channel fading statistics and but also on whether the statistics are known at the receiver and the transmitter. To capture the fading statistics, channel gain is characterized by a single random variable (RV) in SISO systems. But for MIMO systems, the channel is in the form of a matrix, hence the characterization of random matrices plays an indispensable role in studying MIMO channel metrics. 

\par The MIMO system is modelled by an $N_R \times N_T$ channel gain matrix $\vec H$, where $N_R$ is the number of receive antennas and $N_T$ is the number of transmit antennas. Various performance metrics such as capacity, rate, etc. require the eigenvalue statistics of the Gram matrix $\vec{HH}^H$ (or $\vec{HH}^H$). When the elements of $\vec H$ are i.i.d. circular symmetric complex Gaussian with zero or non-zero mean, i.e., when Rayleigh or Rician faded MIMO channels are considered, the Grammian $\vec{HH}^H$ can be characterized by Wishart matrices - central or non-central respectively \cite{james}. These random matrix models have been used widely for deriving capacity expressions in the case of Rayleigh faded MIMO channels \cite{telatar_capacity, chiani_capacity, goldsmith_capacity} and also Rician faded MIMO channels \cite{kang_capacity, mckay_capacity}. Recently, there has been focus on generalized fading models namely $\kappa-\mu$ and $\eta-\mu$ models, which were introduced in the seminal work \cite{yacoub_k_mu}. These distributions model the small scale variations in the fading channel in the line of sight and non-line of sight conditions respectively. Further, these generalized fading distributions include the well-studied Rayleigh, Rician, Nakagami, one-sided Gaussian distributions as special cases. $\kappa-\mu$ and $\eta-\mu$ fading distributions have been widely used in capacity and outage probability analysis for SISO and MISO systems. 

\par Average channel capacity of single branch $\kappa-\mu$ and $\eta-\mu$ faded receivers is studied by the authors of \cite{capacity_costa}. The outage probability (OP), coverage probability and rate of these generalized fading channels are analyzed by works such as \cite{outage_morales, outage_paris, outage_ermolova,suman_coverage, suman_outage} and the references therein. OP of MRC in $\kappa-\mu$ fading channels in the presence of co-channel interference (CCI) is studied by \cite{outage_mrc_morales}. For example, OP expression for $\eta-\mu$ signal of interest (SOI) and Rayleigh faded interferers is derived in terms of confluent Lauricella function in \cite{outage_morales}. OP expressions, when SOI experiences $\eta-\mu$ or $\kappa-\mu$ fading and the interfering signals are subject to $\eta-\mu$ fading, have been derived in \cite{outage_paris}. This was further extended to cases where CCI can be either $\eta-\mu$ or $\kappa-\mu$ fading in \cite{outage_ermolova}. Expressions for coverage probability and rate are derived in terms of Lauricella's function of the fourth kind in \cite{suman_coverage}, when SOI and CCI experience $\kappa-\mu$ and $\eta-\mu$ fading respectively. Approximate OP and rate expressions are derived in terms of the Appell function in \cite{suman_outage}, when the user channel and the interferers experience $\kappa-\mu$ and $\eta-\mu$ fading respectively. OP analysis of $\kappa-\mu$ fading is performed for optimum combining in \cite{srinivasan_optimum}. Secrecy capacity analysis is carried out in \cite{secrecy_bhargav, srinivasan_secrecy} and effective throughput in MISO systems is determined in \cite{throughput_zhang, throughput_zhang1}.  Analysis of decode and forward relay system for generalized fading models is performed in \cite{df_li, df_fikadu, df_kumar}. Asymptotic analysis of generalized fading channels using extreme value theory is performed in \cite{athira_evt}.

\par On the other hand, the capacity of MIMO systems for these generalized fading channels has been less analyzed for want of a random matrix model that characterizes the channel matrix. Nevertheless, some random matrix models have been developed for Nakagami and Rician-shadowed fading channels. A random matrix model has been developed for Nakagami-q fading in \cite{kumar_random} and the pdf of eigenvalues of the Gram $\vec{HH}^H$ is obtained in terms of a Pfaffian. In \cite{gholizadeh_capacity}, the ergodic capacity of MIMO correlated Nakagami-m fading channel has been derived using the concept of a copula. But the work presents an analysis only for $2 \times 2$ MIMO channel and determining the capacity of MIMO channels with a larger number of receive and transmit antennas using this method is cumbersome. Recently a MIMO capacity upper bound was derived in \cite{mimo_vergara} for the $\kappa-\mu$ and $\eta-\mu$ fading channels. In \cite{alfano_mimo,pozas_mimo} a MIMO model has been developed for Rician-shadowed fading as a unification model for MIMO-Rayleigh and MIMO-Rician fading models. But, to the best of our knowledge, no work has presented even an approximate matrix model for the $\kappa-\mu$ and $\eta-\mu$ fading channels.

\par Given the complicated pdf structure of complex variable $\kappa-\mu$ and $\eta-\mu$ fading distributions \cite{k_mu_phase, eta_mu_phase}, it is challenging to develop the matrix distribution and the eigenvalue statistics for $\vec{HH}^H$, even when the elements of $\vec H$ are assumed to be i.i.d. complex $\kappa-\mu$ or $\eta-\mu$ random variables. Hence, in this work, we develop an approximate matrix model for $\vec{HH}^H$ (or $\vec{H}^H\vec{H}$) in terms of a Wishart distribution, which is a very well-studied matrix distribution \cite{james}. There is some prior literature which focuses on approximation of random matrices with Wishart matrices. Approximating any matrix distributions by central Wishart by means of Taylor expansion is studied in \cite{kollo_wishart}, but the approximation requires the knowledge of not only one or more cumulants and moments of the random matrix but also the derivatives of central Wishart matrix. Also, the approximation of non-central Wishart matrix by a central Wishart by means of Laguerre polynomial expansion is given in \cite{tan_wishart} and by means of the moment generating functions in \cite{steyn}. But, the Laguerre polynomial expansion and the moment generating functions are difficult to derive for matrix variate $\kappa-\mu$ and $\eta-\mu$ fading distributions. 

\par Our contribution in this paper, are as follows:
\begin{itemize}
\item[a)]For both complex $\kappa-\mu$ and $\eta-\mu$ distribution\footnote{We use the distribution of the in-phase and quadrature components}, we propose a Wishart distributed approximation of $\vec{HH}^H$, such that the approximation has the least KL divergence from the original matrix distribution.
\item[b)]For complex $\kappa-\mu$, we also propose another simple Wishart approximation of $\vec{HH}^H$, such that the approximation has its first moment matched with the original matrix distribution of $\vec{HH}^H$ and the degree of freedom is constrained to be the number of columns of the matrix $\vec H$. This method is similar to the approximation of Rician channels in \cite{steyn}.
\end{itemize} 
The proposed approximation is discussed in Section \ref{approx}. Though the approximations are derived for $\kappa-\mu$ and $\eta-\mu$ fading, the same methodology can be extended to other generalized fading models like $\kappa-\mu$ shadowed fading \cite{pozas_shadowed, cotton_d2d}, $\alpha-\kappa-\mu$ or $\alpha-\eta-\mu$ fading \cite{fraidenraich_alphakappa}, or $\alpha-\kappa-\mu$ shadowed fading \cite{pablo_alphakappamu}, if the joint complex envelope-phase distributions are available. Even if the exact MIMO matrix models are determined in future, the Wishart approximations derived in our paper will still remain significantly simpler. In Section \ref{application}, the utility of the approximation is shown via three applications namely:
\begin{itemize}
\item[a)] Determining the capacity of MIMO systems with i.i.d. $\kappa-\mu$ or $\eta-\mu$ channel gains. 
\item[b)] Determining OP expressions of MIMO-maximum ratio combining (MIMO-MRC).
\item [c)] Determining zero-forcing (ZF) ergodic rate expressions of single-cell uplink MIMO systems in $\kappa-\mu$ or $\eta-\mu$ fading channels for the scenario in which only low-resolution ADCs are deployed at the receiver RF chains.
\end{itemize} 
To the best of our knowledge, ours is the first work to derive even an approximate capacity expression in the presence of $\kappa-\mu$/$\eta-\mu$ MIMO channels. Also, no prior work has given OP expressions for a receiver diversity system employing MIMO-MRC with $\kappa-\mu$ or $\eta-\mu$ fading channels. To show the utility of our expressions in a recent 5G technology, we have derived rate expressions for ZF receiver in massive MIMO scenario. The derived approximations are compared with Monte-Carlo simulations and a close match is found between the theoretical results and simulation results. While we have only shown the utility of the approximation in three applications namely capacity computation and outage probability computation, the approximation can be used in any application which deals with random $\kappa-\mu$/$\eta-\mu$ matrix models.
\par Basic notation: $E_x(.)$ denotes expectation with respect to distribution x. $|\vec X|$ and $det(\vec X)$ denote determinant of a matrix $\vec X$. $etr(\vec X)$ denotes an exponential raised to trace of the matrix $\vec X$.

\section{Generalized fading models}
In this section, we introduce the complex variable pdfs of commonly used generalized fading models.
\subsection{$\eta-\mu$ model}
The $\eta-\mu$ is a fading distribution that represents small scale fading effects in non-line of sight condition. The elements $h_{ij}$ of $\vec H$ are independent and  identical $\eta- \mu$ distributed random variables with density \cite{eta_mu_phase},
\begin{align}\label{nmu}
f_{x y}(x_{ij}, y_{ij}) &= \frac{\mu ^{2 \mu}|x_{ij}y_{ij}|^{2\mu-1}}{\Omega_X^{\mu} \Omega_Y^{\mu} \Gamma^2(\mu)}exp(-\mu\Big(\frac{x_{ij}^2}{\Omega_X}+\frac{y_{ij}^2}{\Omega_Y}\Big))
\end{align}
where $\Omega$ is the power parameter given by $\Omega=2 \sigma^2 \mu$, $\sigma^2$ is the power of the Gaussian variable in each cluster, $\mu$ is the number of clusters. Note, $\Omega_X=(1-\eta)\Omega/2$,  $\Omega_Y=(1+\eta)\Omega/2$, $-1 \leq \eta \leq 1$.

\subsection{$\kappa-\mu$ model}
In $\kappa-\mu$ fading model, the signal is divided into different clusters of waves. The number of clusters is $\mu$ and in each of the clusters, there is a deterministic LOS component with arbitrary power and scattered waves with identical powers. Note, $\kappa$ is the ratio between the total power of the dominant components and the total power of the scattered waves.
Suppose the elements $h_{i,j}= x_{ij}+ j y_{ij}$ of $\vec H$ are i.i.d. $\kappa- \mu$ random variables,  where $x_{ij}$ and $y_{ij}$ are the real and imaginary components respectively, then the  joint distribution is given by \cite{k_mu_phase},
\begin{align}
\nonumber
f_{x y}(x_{ij}, y_{ij}) &= \frac{|x_{ij}y_{ij}|^{\mu/2}}{4\vec \sigma^4 |pq|^{\mu/2-1}}exp(-\frac{(x_{ij}-p)^2+(y_{ij}-q)^2}{2 \vec \sigma^2})\\
& sech(\frac{px_{ij}}{\vec \sigma^2})sech(\frac{qy_{ij}}{\vec \sigma^2})I_{\frac{\mu}{2}-1}( \frac{  |p x_{ij}|}{\vec \sigma^2} )I_{\frac{\mu}{2}-1}( \frac{  |qy_{ij}|}{\vec \sigma^2} ).\label{kmu}
\end{align}
Here $p^2=\sum_{i=1}^{\mu}p_i^2$ and $q^2=\sum_{i=1}^{\mu}q_i^2$, where $p_i$ and $q_i$ are the LOS components of in-phase and quadrature components respectively of multipath waves of each cluster. $\kappa= \frac{p^2+q^2}{2\mu\sigma^2}$, where $\sigma^2$ is the power of the scattered waves.

\subsection{Special case: Nakagami-m model}
If the elements $h_{ij}$ of $\vec H$ are independent and identically distributed Nakagami-m variables with density,
\begin{align*}
f_{xy}(x_{ij},y_{ij})= \frac{m^{m}|x_{ij}|^{m-1} |y_{ij}|^{m-1}}{\Omega^{m}\Gamma^2(m/2)}exp(-\frac{m}{\Omega}(x_{ij}^2+y_{ij}^2)).
\end{align*}
We can obtain a Nakagami  random variable by substituting $\eta=0$ i.e.,  $\Omega_X=\Omega_Y=\Omega/2$ and $m=2 \mu$ in $\eta-\mu$ random variable given by (\ref{nmu}). It can also be obtained by substituting $\kappa=0$ i.e., $p=q=0$ and $\Omega=2\mu \sigma^2$ in $\kappa-\mu$ random variable given by (\ref{kmu}). 

\section{Proposed Matrix Approximations}\label{approx}
Though the statistical characterization of MIMO fading channels is of interest for generalized fading models, it is very extremely challenging to derive the eigenvalue distribution for these models. Hence, it is essential to develop at least an approximate matrix model to study metrics like capacity. In the following subsection, we explain why it is  challenging to develop the matrix distribution and the eigenvalue statistics for $\vec{HH^H}$, when the elements of $\vec H$ are assumed to be i.i.d. complex $\kappa-\mu$ or $\eta-\mu$ random variables. In the subsequent subsections, we describe the proposed approximation in detail. 

\subsection{Why is deriving the exact eigenvalue statistic intractable?}
The eigenvalue statistics of any Grammian $\vec W= \vec {HH}^H$ can be obtained by first decomposing $\vec H$ as $\vec H= \vec {LQ}$. Here, $\vec L$ is a complex triangular matrix with real positive diagonal elements and $\vec Q$ is a complex unitary matrix. Hence, the distribution $p(\vec H)$ is transformed to $p(\vec L, \vec Q)$. Next, integrating over $\vec Q$, we obtain the distribution $p(\vec L)$. Then, we perform the transformation $\vec W= \vec{LL}^H$ and obtain the distribution of the Grammian $p(\vec W)$. Finally, we perform eigenvalue decomposition $\vec W= \vec{S\Lambda S}^H$ and integrate over $\vec S$ to obtain the joint eigenvalue distribution of $\vec W$.  This method used in \cite{fraidenraich_capacity} to determine the eigenvalue statistics of $2 \times 2$ $\vec W= \vec {HH}^H$, where the elements $h_{ij}= x_{ij}+j y_{ij}$ are i.i.d. complex Nakagami-m random variables with uniform phase. 
\par Consider that the elements $h_{ij}= x_{ij}+j y_{ij}$ are i.i.d. complex $\eta-\mu$ random variables with pdf (\ref{nmu}). The joint distribution is simply the product of the pdf of the individual elements. Therefore, the joint distribution of elements of a $2 \times 2$ matrix $\vec H$ for the complex $\eta-\mu$ case is given by
\begin{align}
p(\vec H)= K exp(-\frac{\mu tr(\vec{HH}^H)}{\Omega_X}) \prod_{i,j=1}^2 |x_{ij}y_{ij}|^{2\mu-1} exp(-\mu y_{ij}^2 \Big(\frac{1}{\Omega_X}-\frac{1}{\Omega_Y}\Big)),
\end{align}
where $K= (\frac{\mu ^{2 \mu}}{\Omega_X^{\mu} \Omega_Y^{\mu} \Gamma^2(\mu)})^4$. 
Using LQ decomposition, the matrix $\vec H$ can be decomposed as $\vec H= \vec{LQ}$, where the matrix $
\vec Q$ is given by \cite{fraidenraich_capacity}
\begin{align}
\vec Q= \begin{pmatrix}
e^{j \phi_1} cos \theta & e^{j \phi_2}sin \theta\\
-e^{j(\phi_3-\phi_2)}sin \theta & e^{j(\phi_3-\phi_1)} cos \theta
\end{pmatrix}
\end{align}
with $0 \leq \phi_1, \phi_2, \phi_3 \leq 2\pi$, $0 \leq \theta \leq \pi/2$. The matrix $\vec L$ is given by
\begin{align}
\vec L= \begin{pmatrix}
l_{11} & 0\\
l_{21R}+ j l_{21I} & l_{22}
\end{pmatrix}
\end{align}
where $l_{11}$ and $l_{22}$ are real. From the Jacobian of this transformation given by $|J|= l_{11}^3l_{22} sin \theta cos \theta$, the joint pdf of $\vec L$ and $\vec Q$ is obtained. In the case of Rayleigh distribution, $p(\vec H)$ doesn't depend on $\vec Q$, hence there is no integration involved over elements of $\vec Q$. Integrating over the elements of $\vec Q$, namely $\phi_1, \phi_2, \phi_3$ and $\theta$, for the case of $\eta-\mu$ distribution becomes intractable. A similar case is seen for $\kappa-\mu$ fading also. This methodology leads to a simpler form only for Nakagami-m RV with uniform phase \cite{fraidenraich_capacity}. Even in the case of Nakagami-m RV with a non-uniform phase, the eigenvalue distribution obtained by the above methodology is not in closed form \cite{gholizadeh_capacity}. Summarizing, the above method cannot be adopted for complex $\eta-\mu$/$\kappa-\mu$ scenario.

\subsection{Minimizing K-L divergence}
An alternative to determining the exact matrix distribution is finding a Wishart distribution that has the least K-L divergence from the actual distribution. As seen earlier, Wishart distribution is a well-studied distribution and its properties can be readily applied \cite{james}. Let $\vec H$ be an $n_1 \times n_2$ random matrix with independent and identically distributed elements and $\vec X= \vec H \vec H^H$ be an $ n_1 \times n_1$ random matrix. Say, the exact matrix distribution of $\vec X$ denoted by $p(\vec X)$ is not known, as in the case of $\eta-\mu$/$\kappa-\mu$/Nakagami-m distributions. Let $q(\vec X)$ be that Wishart distribution which minimizes the K-L divergence between $p(\vec X)$ and all the complex Wishart distributions $ \mathcal{CW}_{n_1}(n, \vec \Sigma)$, i.e., 
\begin{align}\label{minimizer}
\nonumber
q(\vec X) = \underset{ q(\vec X)}{argmin} KL( p(\vec X) || q(\vec X)) &=\underset{ q(\vec X)}{argmax} \int p(\vec X)[ln (q (\vec X))-ln (p(\vec X)) ]d\vec X\\
&= \underset{ q(\vec X)}{argmax} \int p(\vec X)ln (q (\vec X)) d\vec X. 
\end{align}
Note that, we assume an unknown degrees of freedom as $n$ in this case, while in the Wishart approximation we had constrained the degree of freedom to $n_2$.
The density of an $n_1 \times n_1$ complex Wishart matrix $\vec X \sim \mathcal{CW}_{n_1}(n, \vec \Sigma)$ is given by \cite{randombook},
\begin{align*}
q(\vec X)= \frac{1}{C\Gamma_{n_1}(n) (det \vec \Sigma)^n} etr(-\vec \Sigma^{-1} \vec X) (det \vec X)^{n-n_1},
\end{align*}
where $C\Gamma_.(.)$ is the complex multivariate gamma function \cite{randombook}.
Substituting the density in (\ref{minimizer}), we obtain
\begin{align*}
q(\vec X)
&= \underset{ q(\vec X)}{argmax} \int p(\vec X) \bigg[-ln (C\Gamma_{n_1}(n)) -n ln|\vec \Sigma| + Tr(-\vec \Sigma^{-1} \vec X) + (n-n_1) ln |\vec X| \bigg] d\vec X\\
&= \underset{ q(\vec X)}{argmax} \bigg[-ln (C\Gamma_{n_1}(n)) -n ln|\vec \Sigma|  + Tr(-\vec \Sigma^{-1} E_{p}[\vec X]) + (n-n_1) E_{p}[ln |\vec X|] \bigg].
\end{align*}
Denoting $\vec Z= E_{p}[\vec X]$ and $Y= E_{p}[ln |\vec X|]$, we get
\begin{align}\label{qequation}
q(\vec X) &=  \underset{ q(\vec X)}{argmax} \bigg[-ln (C\Gamma_{n_1}(n)) -n ln|\vec \Sigma| + Tr(-\vec \Sigma^{-1} \vec Z) + (n-n_1) Y \bigg].
\end{align}
To obtain the minimizing distribution, we can differentiate the above equation with respect to two variables namely, $\vec \Sigma$ and $n$. Differentiating equation (\ref{qequation}) w.r.t. $\vec \Sigma$, we obtain
\begin{equation*}
\frac{dq(\vec X)}{d \vec \Sigma}= -n \vec \Sigma^{-1} + \vec \Sigma^{-1} \vec Z^{T} \vec \Sigma^{-1}.
\end{equation*}
When the above equation is equated to zero, we obtain 
\begin{align}\label{sigma1}
 \vec \Sigma = \frac{1}{n} \vec Z^{T}= \frac{1}{n} E_{p}[\vec X].
\end{align}
Note that we obtain the same $\vec \Sigma$ when we equate the expectations of the matrix with respect to distributions $p(\vec X)$ and $q(\vec X)$, i.e.,  $E_{p}[\vec X]=E_{q}[\vec X]$. 
Now differentiating equation  (\ref{qequation}) w.r.t. $n$, we obtain
\begin{align*}
\frac{dq(\vec X)}{d n} &=- ln |\vec \Sigma| + Y - \sum_{i=1}^{n_1} \psi(n-i+1),
\end{align*}
where $\psi(.)$ is the digamma function \cite{granstrom}.
Equating the derivative to zero, we get
\begin{align*}
- ln |\vec \Sigma| + Y - \sum_{i=1}^{n_1} \psi(n-i+1)=0.
\end{align*}
By substituting  $\vec \Sigma = \frac{1}{n} \vec Z^{T}$ from (\ref{sigma1}), we obtain,
\begin{align}\label{nequation}
n_1ln(n) - ln |\vec Z| + Y - \sum_{i=1}^{n_1} \psi(n-i+1)=0.
\end{align}
Matching the expectations $E_{p}[ln |\vec X|] = E_{q}[ln |\vec X|]$ also leads to (\ref{nequation}).
\par Hence minimizing the K-L divergence has reduced to a simple case of matching the expectations $E_p[\vec X]$ and $ E_p[ln |\vec X|]$ with $E_q[\vec X]$ and $ E_q[ln |\vec X|]$ respectively. Though this is the most ideal approximation, solving (\ref{nequation}) for $n$ requires the knowledge of $Y=E_p[ln |\vec X|]$. Since finding the actual random matrix variate distribution of $\vec X$ is intractable, finding the exact expectation of the log-determinant $E_p[ln |\vec X|]$ analytically is not possible. Nevertheless, it is possible to approximate the log determinant $ln |\vec X|$ \cite[Page 644]{boyd2004convex}. One of the simplest approximations is a second order Taylor expansion of the $ln |\vec X|$ at $E_p[\vec X]$ given by,
\begin{equation}
    ln |\vec X| \approx ln|E_p[\vec X]|+ tr\left( E_p[\vec X]^{-1}\left(\vec X- E_p[\vec X]\right)\right) -\frac{1}{2} tr\left( E_p[\vec X]^{-1}\left(\vec X- E_p[\vec X]\right)E_p[\vec X]^{-1}\left(\vec X- E_p[\vec X]\right)\right).
\end{equation}
Evaluating the expectations on both sides, we obtain
\begin{equation}\label{Eplnx}
    E_p[ln |\vec X|] \approx ln|E_p[\vec X]| -\frac{1}{2}tr E_p\left( E_p[\vec X]^{-1}\vec X E_p[\vec X]^{-1}\vec X\right) +\frac{1}{2}n_1.
\end{equation}
The above approximation can now evaluated for $\eta-\mu$ and $\kappa-\mu$ fading. 
\subsubsection{$\eta-\mu$} 
Note that the approximation involves the mean of the random matrix given by $\vec Z=E_p[\vec X]=E_p[\vec{HH}^H]$. The diagonal elements $z_{ii}$ are means of sums of $\eta-\mu$ envelope square variables.
i.e., $z_{ii}=  E[\sum_{j=1}^{n_2}[x_{ij}^2 + y_{ij}^2]]$. Hence by \cite{yacoub_k_mu}, $z_{ii}=n_2 (\Omega_X + \Omega_Y)$. The off-diagonal elements are given by,
\begin{align*}
z_{ij} &=   E[ \sum_{k=1}^{n_2}[ (x_{ik}+j y_{ik})(x_{kj}-j y_{kj})]].
\end{align*}
Since $x_{ik}$ and $y_{ik}$ are i.i.d., we obtain $\forall i,j$ and $i \neq j$,
\begin{align*}
z_{ij}&=\sum_{k=1}^{n_2}(  E[ x_{ik}]^2 +   E[y_{kj} ]^2) =0.
\end{align*}
The off-diagonal elements are zero, because distributions $f_x(x_{ij})$ and $f_y(y_{ij})$ are odd functions.
Therefore,  $\vec Z= E_p[\vec X]=n_2(\Omega_X + \Omega_Y)\vec  I_{n_1}$. Also, 
$ln|E_p[\vec X]|= n_1ln\left(n_2(\Omega_X + \Omega_Y)\right)$. Also, $tr E_p\left( E_p[\vec X]^{-1}\vec X E_p[\vec X]^{-1}\vec X\right)$ evaluates to
\begin{align*}
    tr E_p\left( E_p[\vec X]^{-1}\vec X E_p[\vec X]^{-1}\vec X\right) &= \left(n_2(\Omega_X + \Omega_Y)\right)^{-2}tr E_p\left( \vec X \vec X\right)\\
   &= \frac{n_2 n_1}{n_2\left(\Omega_X + \Omega_Y\right)^{-2}} \Big((\frac{\mu+1}{\mu}+n_2-1)(\Omega_X^2+\Omega_Y^2) \\
   & \quad + 2n_2\Omega_X \Omega_Y + (n_1-1)(\Omega_X + \Omega_Y)^2\Big)
\end{align*}
Hence,
\begin{align*}
     E_p[ln |\vec X|] &\approx n_1ln\left(n_2(\Omega_X + \Omega_Y)\right)-\frac{0.5n_2 n_1}{n_2\left(\Omega_X + \Omega_Y\right)^{-2}} \Big((\frac{\mu+1}{\mu}+n_2-1)(\Omega_X^2+\Omega_Y^2) \\
   & \quad + 2n_2\Omega_X \Omega_Y + (n_1-1)(\Omega_X + \Omega_Y)^2\Big) +\frac{1}{2}n_1
\end{align*}
Once $E_p[\vec X]$ and $ E_p[ln |\vec X|]$ are obtained, the $\vec \Sigma$ and $n$ of the minimizer-Wishart distribution are obtained by (\ref{sigma1}) and (\ref{nequation}). For $\eta-\mu$ faded channels, $\vec \Sigma$ will be diagonal.
\subsubsection{$\kappa-\mu$ model}
The diagonal elements of $\vec Z= E_p[\vec X]=E_p[\vec{HH}^H]$, i.e, $z_{ii}$ are nothing but the mean of $n_2$ $\kappa-\mu$ envelope square variables.
i.e., $z_{diag}=z_{ii}=  E[\sum_{j=1}^{n_2}[x_{ij}^2 + y_{ij}^2]]$. Hence,
\begin{equation}\label{zdiag}
    z_{diag}=z_{ii}=2 \sigma^2 n_2 (1+\kappa) \mu,
\end{equation} 
where $\kappa= \frac{p^2+q^2}{2 \mu \sigma^2}$ \cite{yacoub_k_mu}. 
The off diagonal elements of $z_{ij}$ are given by
\begin{align}
\nonumber
z_{off}=z_{ij} &=   E[ \sum_{k=1}^{n_2}[ (x_{ik}+j y_{ik})(x_{kj}-j y_{kj})]]= \sum_{k=1}^{n_2}  E[ x_{ik}x_{kj} +y_{ik} y_{kj} -jx_{ik}y_{kj} +j x_{kj}y_{ik} ].
\end{align}
Since $x_{ik}$ and $y_{ik}$ are i.i.d., we obtain $\forall i,j$,
\begin{align}\label{zij}
z_{off}=z_{ij} &=\sum_{k=1}^{n_2}((  E[ x_{ik}])^2 + (  E[y_{kj} ])^2).
\end{align}
$  E[x_{ik}]$ and $  E[y_{ik}]$ are given by,
\begin{align}\label{exint}
  E[x_{ik}]= \int_{- \infty}^{\infty} x \frac{|x|^{\mu/2}}{2\vec \sigma^2 |p|^{\mu/2-1}}exp(-\frac{(x-p)^2}{2 \vec \sigma^2})sech(\frac{px}{\vec \sigma^2})I_{\frac{\mu}{2}-1}( \frac{  |p x|}{\vec \sigma^2} )dx
\end{align}
\begin{align}\label{eyint}
  E[y_{ik}]= \int_{- \infty}^{\infty} y \frac{|y|^{\mu/2}}{2\vec \sigma^2 |q|^{\mu/2-1}}exp(-\frac{(y-q)^2}{2 \vec \sigma^2})sech(\frac{qy}{\vec \sigma^2})I_{\frac{\mu}{2}-1}( \frac{  |qy|}{\vec \sigma^2} )dy
\end{align}
$\forall i,j$.
 The closed form expressions for the above integrals seem mathematically intractable. However an approximation for the integral is derived in Appendix \ref{off_kmu} and given by (\ref{xapprox}) and (\ref{yapprox}). \footnote{These closed form approximations are also compared with both numerical integration evaluation of (\ref{exint}) and (\ref{eyint}) and Monte-Carlo simulation and an excellent match is observed with both.}
Since all the $\kappa-\mu$ elements of the matrix $\vec H$ are i.i.d., the mean of all the off-diagonal elements are equal.  Substituting the results from (\ref{xapprox}) and (\ref{yapprox}) in (\ref{zij}), we obtain $\forall i,j$ and $i \neq j$,
\begin{align}\label{zijapprox1}
\nonumber
z_{ij} & \approx n_2 \Bigg[\Big[2pe^{-\frac{p^2}{2  \sigma^2}}(\frac{4 \sigma^2}{4 \sigma^2+2p^2 \pi})^{\mu/2 +1}\frac{\Gamma(\mu/2+1)}{\Gamma(\mu/2)}
 \Psi_1( \mu/2+1,1,3/2, \mu/2,\frac{2 p^2 \pi}{2 p^2 \pi+4 \sigma^2},\frac{2 p^2}{ 4 \sigma^2+ 2p^2 \pi})\Big]^2\\
 & \quad +  \Big[2qe^{-\frac{q^2}{2  \sigma^2}}(\frac{4 \sigma^2}{4 \sigma^2+2q^2 \pi})^{\mu/2 +1}\frac{\Gamma(\mu/2+1)}{\Gamma(\mu/2)}
 \Psi_1( \mu/2+1,1,3/2, \mu/2,\frac{2 q^2 \pi}{2 q^2 \pi+4 \sigma^2}, \frac{2 q^2}{ 4 \sigma^2+ 2q^2 \pi})\Big]^2\Bigg].
\end{align}
Now, to evaluate (\ref{Eplnx}), note that
\begin{align}
    ln|E_p[X]|=(z_{diag}-z_{off})+ n_1 z_{off}
\end{align}
Similarly, we need to determine
$tr E_p\left( E_p[\vec X]^{-1}\vec X E_p[\vec X]^{-1}\vec X\right)$.
It is easy to derive the closed form expression for $\vec X$ with dimension $2 \times 2$, i.e., when either $N_T$ or $N_T$ can be fixed at $2$. This is shown in Appendix \ref{2by2}. Once $E_p[\vec X]$ and $ E_p[ln |\vec X|]$ are obtained, the $\vec \Sigma$ and $n$ of the minimizer-Wishart distribution are obtained by (\ref{sigma1}) and (\ref{nequation}). Unlike $\eta-\mu$ fading, we do not obtain a diagonal covariance matrix $\vec \Sigma$ for $\kappa-\mu$ fading. The derivation becomes algebraically complicated with increasing $N_R$ or $N_T$, but not intractable. To circumvent even this algebraic complexity, we also come up with an alternative approximation, which retains matching the first moment.

\subsection{1-Moment matching approximation for $\kappa-\mu$ fading}
We approximate the density $p(\vec X)$ by an $n_1 \times n_1$ Wishart matrix whose distribution is $q(\vec X) =\mathcal{CW}_{n_1}(n, \vec \Sigma)$  with $n$ degrees of freedom and covariance matrix $\vec \Sigma$, such that $E_{p}[\vec X]=E_{q}[\vec X]$, i.e., their first moments are matched
\begin{align}\label{sigma}
E_q(\vec X)= n_2 \vec \Sigma= E_{p}[\vec X]
\end{align}
and the degrees of freedom is fixed to be equal to the number of columns of $\vec H$ i.e., $n=n_2$. This approximation is inspired from the popular paper \cite{steyn}, where a non-central Wishart matrix $\mathcal{CW}_{n_1}(n_2, \vec \Sigma', \vec{MM}^H)$ is approximated by a Wishart matrix $\mathcal{CW}_{n_1}(n_2, \vec \Sigma)$, where $\vec \Sigma= \vec \Sigma' + \frac{1}{n_2} \vec{MM}^H$, by first moment matching. Note, $n_2$ denotes the number of transmitter antennas or the number of interferers in the MIMO channel matrix. For any Wishart distributed matrix $\vec A = \vec{BB}^H \sim  \mathcal{CW}_{n_1}(n_2, \vec \Sigma)$, the degrees of freedom also denote the number of columns of complex Gaussian $\vec B$. Hence, we retain the same number $n_2$, even in approximation, given that there is no correlation in the transmitter side and all the elements of the matrix are i.i.d. The covariance matrix of the corresponding Wishart matrix is given by  $\vec \Sigma = \frac{1}{n_2} (  E_{p}[\vec X])^T=  \frac{1}{n_2} \vec Z^T$. 
Therefore, for $\eta-\mu$ fading,  $\vec \Sigma= (\Omega_X + \Omega_Y)\vec  I_{n_1}$. It is interesting to note that, the approximation doesn't depend on $\eta$. Further, it doesn't depend on $\mu$ if we normalize the powers. Hence, this approximation is not useful in characterizing the $\eta$ and $\mu$ dependence of the $\eta-\mu$ MIMO matrix. On the other hand, for $\kappa-\mu$ fading, since $\vec \Sigma= \frac{\vec Z}{n_2}$, we have
\begin{equation}
\vec \Sigma_{ii} =z_{diag}/n_2=2 \sigma^2 (1+\kappa) \mu \: \text{and} \:
\vec \Sigma_{ij} = z_{off}/n_2=z_{ij}/n_2, \, i \neq j. \label{kmusigma}
\end{equation}
Though this approximation does not minimize the KL divergence, it is much easier to evaluate in case of $\kappa-\mu$ fading.

\subsection{What about more generalized fading models?}
Recently there has been significant focus on more generalized fading models such as $\kappa-\mu$ shadowed fading \cite{pozas_shadowed, cotton_d2d}, $\alpha-\kappa-\mu$ fading, $\alpha-\eta-\mu$ fading \cite{fraidenraich_alphakappa}, $\alpha-\kappa-\mu$ shadowed fading \cite{pablo_alphakappamu}, etc.\footnote{To the best of our knowledge even the phase distribution of these fading models have not been derived in literature.} Once $E_p[\vec X]$ and $ E_p[ln |\vec X|]$ are obtained for these generalized fading models from the complex phase-envelope distribution, the $\vec \Sigma$ and $n$ of the minimizer-Wishart distribution can be obtained by (\ref{sigma1}) and (\ref{nequation}). For all these cases, there are currently no exact MIMO model available in literature. Even if they are determined in near future, the Wishart approximation will still be a simpler approximation and easily amenable to analysis. \footnote{For example, even the exact MIMO model derived in \cite{gholizadeh_capacity} for the well-studied Nakagami-m fading channel with non-uniform phase distribution is not in closed form.}

\section{Applications and Numerical results}\label{application}
In this section, to demonstrate the utility of our work, we apply the above approximation in three very different applications. We first determine the MIMO channel capacity for $\kappa-\mu$/$\eta-\mu$ faded channel coefficients. We then determine the outage probability of MIMO-MRC for $\kappa-\mu$/$\eta-\mu$ faded channel coefficients. Finally, we determine the ZF ergodic rate expressions of massive MIMO employing ADCs. Finding ergodic MIMO channel capacity involves finding the expectation of log determinant of Gram matrix $\vec{HH}^H$, where entries of $\vec H$ are $\kappa-\mu$/$\eta-\mu$ faded. On the other hand, finding OP expressions for MIMO-MRC involves characterizing the CDF of the maximum eigenvalue of $\vec{HH}^H$. Determining ergodic rate expressions involves determining the expectation of the inverse of $\vec{HH}^H$.
 
\subsection{MIMO channel Capacity}
We consider an $N_R \times N_T$ MIMO channel matrix $\vec H$, where $N_R$ denotes the number of receive antennas and $N_T$ denotes the number of transmit antennas. Let $\vec x$ be the $N_T \times 1$ transmitted vector and $\vec n$ be the $N_R \times 1$ zero mean i.i.d. complex Gaussian noise vector. The $N_R \times 1$ received vector $\vec y$ is given by, $\vec y= \vec H \vec x + \vec n$.
Assuming that the transmitter has no channel state information (CSI), the capacity of the MIMO channels when the transmitter has no CSI, is given by\cite{chiani_capacity},
\begin{align}\label{c'}
C'= log_2 \, det( \vec I + \frac{\rho}{N_T} \vec {HH}^H),
\end{align}
where $\rho$ is the average signal to noise ratio (SNR) per receiving antenna.\footnote{Though, for capacity, expectation of $log_2 \, det( \vec I + \frac{\rho}{N_T} \vec {HH}^H)$ can be matched instead of that of $log_2 \, det(\vec {HH}^H)$(as done in Section III), the resultant Wishart matrix will not be a KL divergence minimizer}.
Since $\vec{HH}^H$ and $\vec {H^HH}$ have the same non-zero eigenvalue statistics, from \cite{chiani_capacity},
\begin{align*}
C'= \sum_{i=1}^{n_1} log_2(1+ \frac{\rho}{N_T} \lambda_i),
\end{align*}
where $n_1=min(N_R, N_T)$ and $\lambda_1,...., \lambda_{n_1}$  are the non-zero eigenvalues of $\vec R$, which is given by,
\begin{align*}
\vec R= \begin{cases}
\vec{HH}^H  \quad if \: N_R  \leq N_T\\
\vec{H}^H \vec H  \quad if \: N_R > N_T.
\end{cases}
\end{align*}
Hence, the mean value of $C'$ is given by \cite{telatar_capacity},
\begin{align*}
C= E_{\vec \Lambda}\Big[\sum_{i=1}^{n_1} log_2(1+ \frac{\rho}{N_T} \lambda_i)\Big].
\end{align*}
We do not know the exact eigenvalue distribution of $\vec R$, when $\vec H$ comprises i.i.d. $\kappa-\mu$ or $\eta-\mu$ variables. Hence, we apply the Wishart approximation developed in the last section and then determine $C$. 
\subsubsection{$\eta-\mu$}
For $\eta-\mu$ faded channel, we approximate the $N_R \times N_R$ matrix $\vec{HH}^H$ by a Wishart matrix $\mathcal{CW}_{N_R}(n, b \vec I_{N_R})$, where $b$ and $n$ are obtained by (\ref{sigma1}) and (\ref{nequation}). Once the Wishart approximation is obtained, we can draw expressions from the vast literature of expressions derived for Rayleigh faded MIMO channels. For example, the expressions given in \cite{telatar_capacity} can be used directly as follows:
\begin{equation}\label{etamucapacity}
C \approx \sum_{k=0}^{N_R-1} \int_{0}^{\infty} log_2(1+ \frac{\rho b}{N_T} \lambda) \lambda^{n-N_R} e^{-\lambda} ( L_{k}^{n-N_R}(\lambda))^2 d \lambda
\end{equation}
The above expression is valid for $N_R < N_T$. In case $N_T < N_R$, we approximate $\vec {H}^H \vec H$ instead of $\vec{HH}^H$ since both have the same non-zero eigenvalues. 
\begin{figure}
\centering
\subcaptionbox{ Capacity vs SNR for $N_T=2$, $\mu=2$ and $\eta=0.1$ \label{capacityfignmu1}}{\includegraphics[scale=0.4]{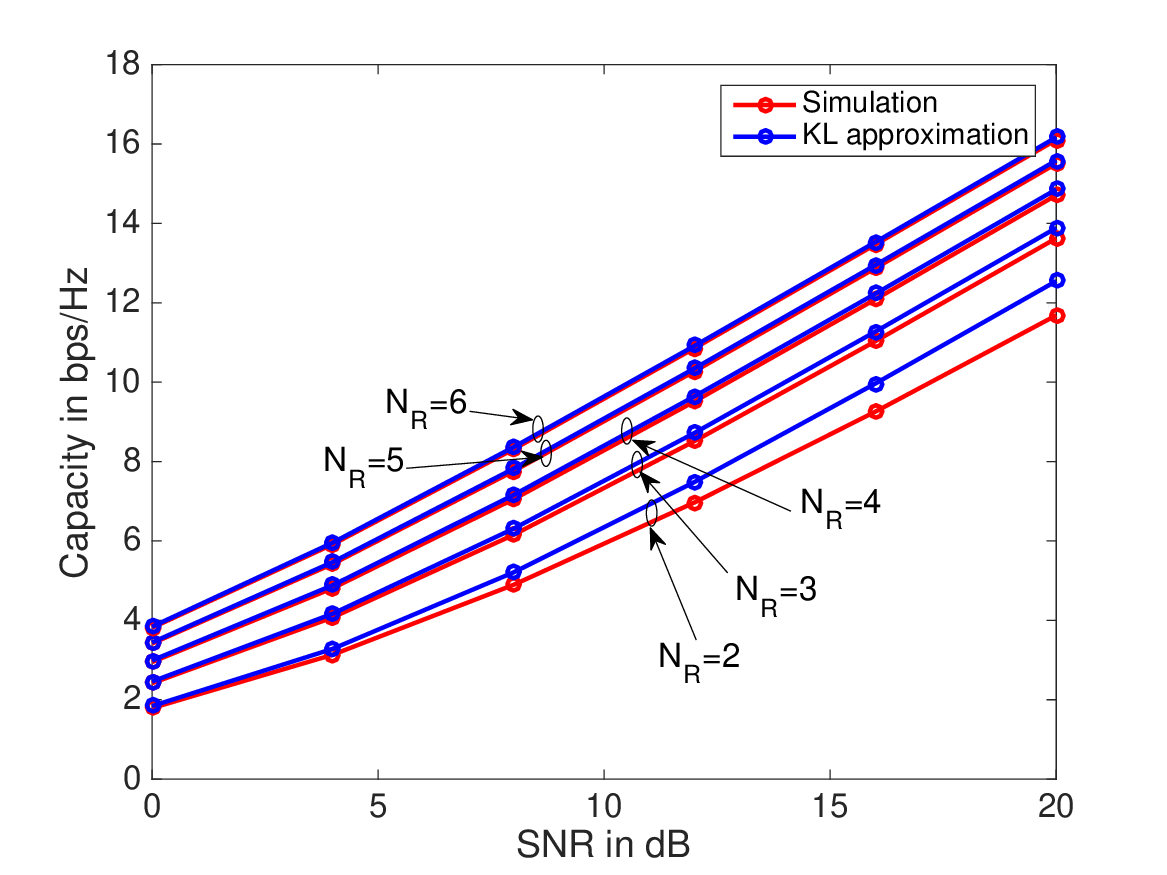}}
\hfill
\subcaptionbox{Capacity vs SNR for $N_R=2$ and $\mu=1$ \label{capacityfignmu2}}{\includegraphics[scale=0.35]{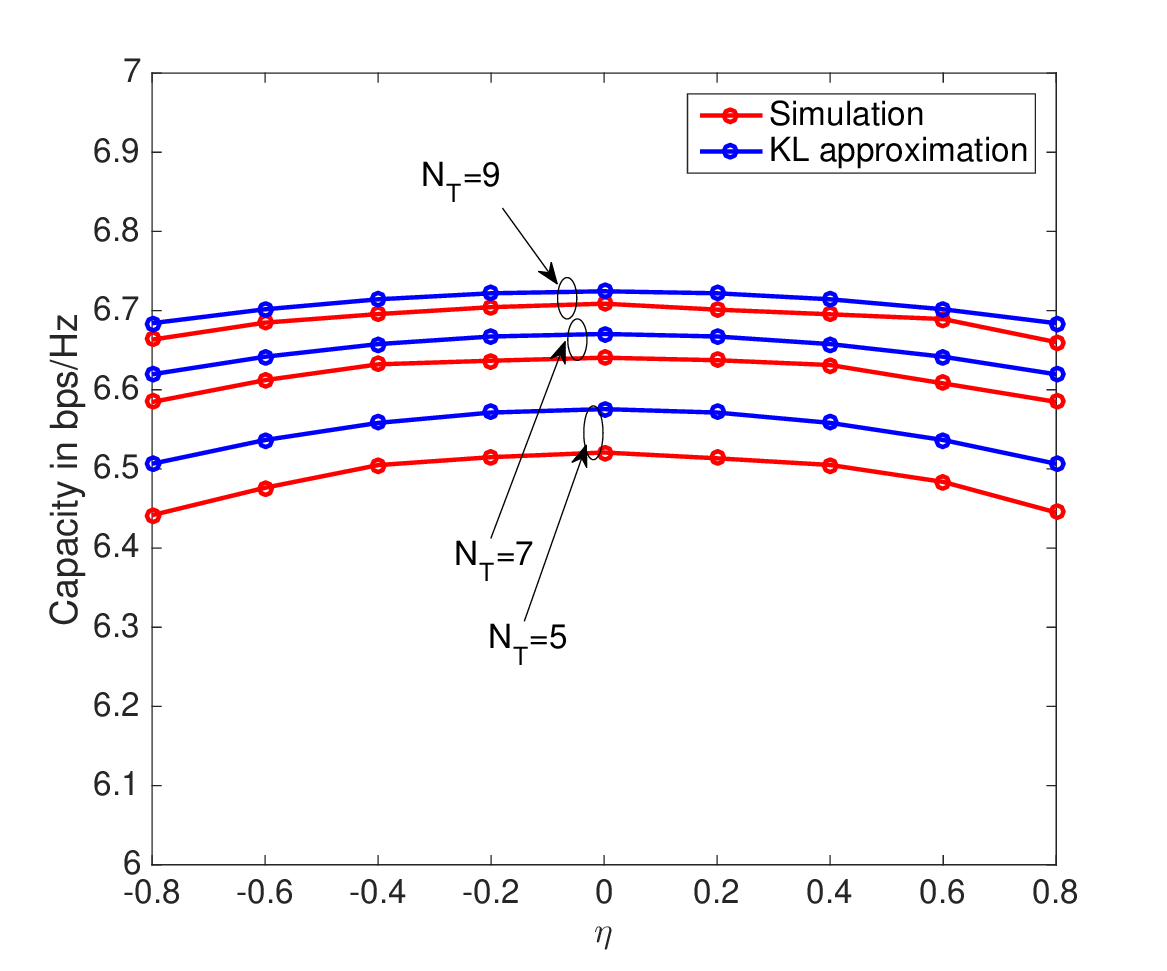}}
\caption{Capacity for varying $\eta$, $\mu$, $N_R$ and $N_T$}
\label{capacityfig1}
\end{figure}
\subsubsection{$\kappa-\mu$}
Unlike $\eta-\mu$ fading, $\kappa-\mu$ fading approximation involves a non-diagonal covariance matrix with repeated eigenvalues. While a lot of results exists for diagonal covariance matrix, the results for non-diagonal covariance matrix is limited, especially when the covariance matrix has repeated eigenvalues. Hence, the approximate expression for $C$ is derived in Appendix \ref{capacity}. For $N_T  \geq N_R$, by substituting $n_1=N_R$, $w_1=2\sigma^2(1+\kappa) \mu-y$ and $w_2=  2 \sigma^2  (1+\kappa) \mu +(N_R-1)y$ which are the eigenvalues of $\vec \Sigma^{-1}$ in (\ref{kmu_capacity}), we can get the average capacity approximation. $\vec \Sigma$ and $n_2=n$ of the Wishart matrix are obtained by (\ref{sigma1}) and (\ref{nequation}). In case $N_T < N_R$, we approximate $\vec {H}^H \vec H$ instead of $\vec{HH}^H$ since both have the same non-zero eigenvalues. For $N_T  \leq N_R$, we substitute $n_1=N_T$ in (\ref{kmu_capacity}) to get the capacity approximation. Here, $w_1=2\sigma^2(1+\kappa) \mu-y$ and $w_2=  2 \sigma^2  (1+\kappa) \mu +(N_T-1)y$ are the eigenvalues of $\vec \Sigma$ with multiplicity $N_T-1$ and $1$ respectively. 
\begin{figure}
\centering
\subcaptionbox{ Capacity vs SNR for $N_R=3$ and $N_T=4$ \label{capacityfigkmu1}}{\includegraphics[scale=0.4]{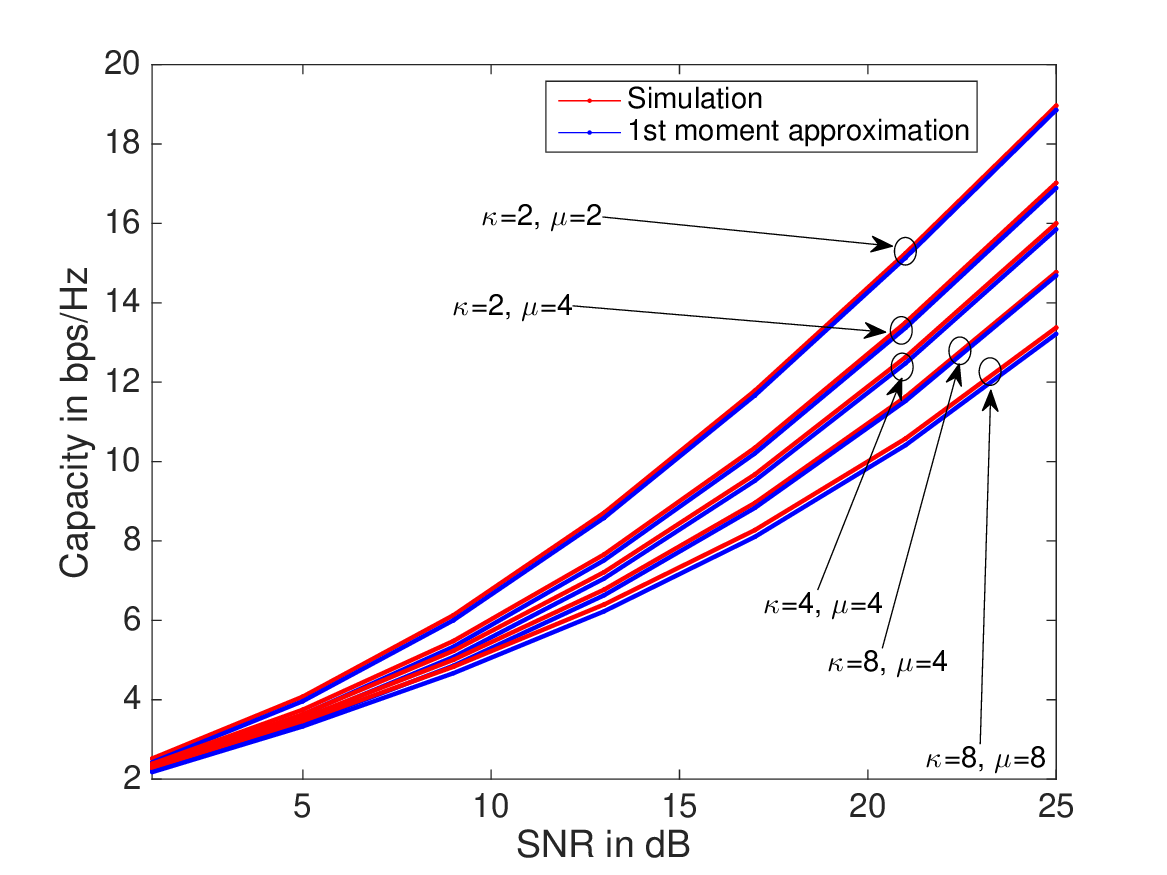}}
\hfill
\subcaptionbox{Capacity vs SNR for $\kappa=2$ and $\mu=2$ \label{capacityfigkmu2}}{\includegraphics[scale=0.4]{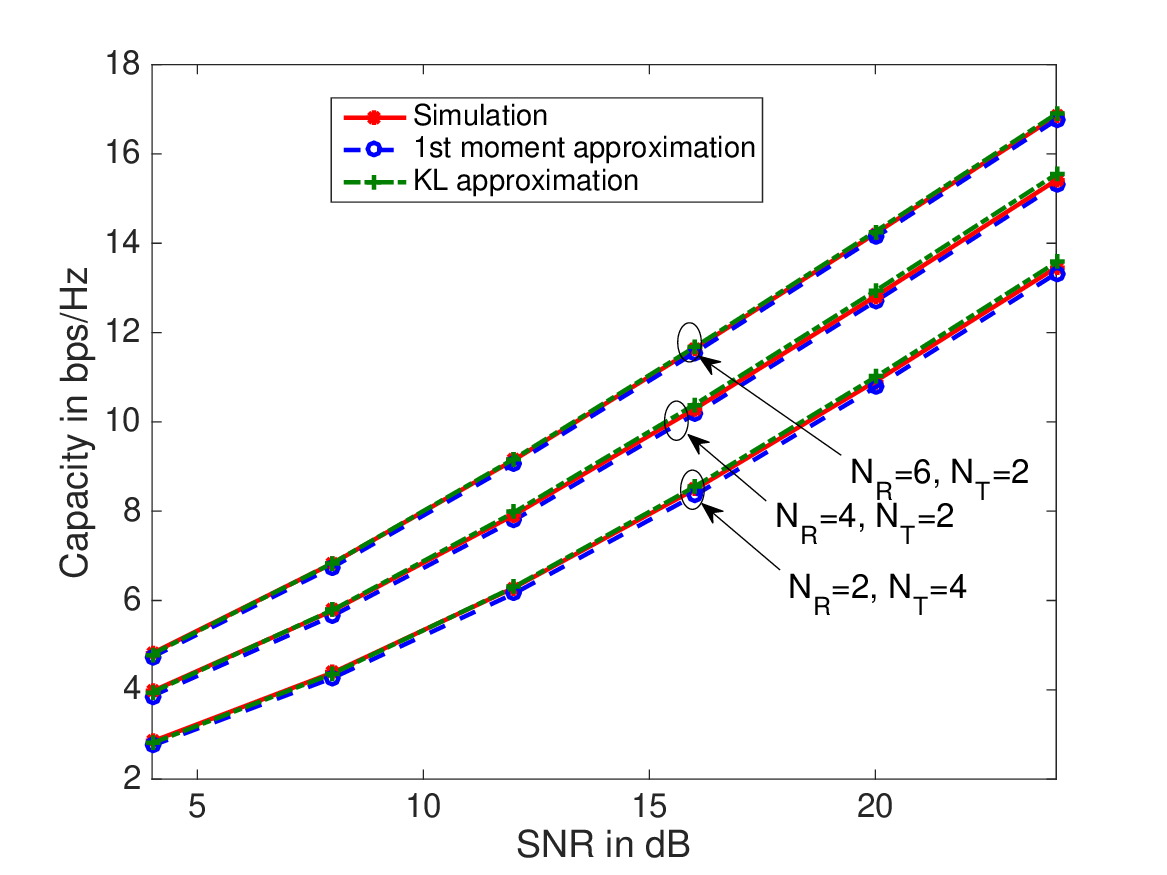}}
\caption{Capacity for varying $\kappa$, $\mu$, $N_R$ and $N_T$}
\label{capacityfig2}
\end{figure}

The derived capacity expressions are verified using Monte-Carlo simulations. For each Monte-Carlo simulation, the $N_R \times N_R$ random matrix $ \vec{HH}^H$ is generated such that $\vec H$ has i.i.d. $\kappa-\mu$ or $\eta-\mu$ complex variables following the distribution that is given in \cite{k_mu_phase, eta_mu_phase}. For a given SNR $\rho$ and $N_T$, capacity is evaluated using (\ref{c'}). This procedure is repeated over many realizations of $\vec{HH}^H$ and the mean is taken to obtain the average capacity. This procedure is repeated for various values of $\kappa$/$\eta$, $\mu$, $N_R$ and $N_T$. A close match is found between the theoretical and simulation results for all the cases as can be seen from the Fig. \ref{capacityfig1}- Fig. \ref{capacityfig2}. For the case of $\eta-\mu$ distribution, from Fig. \ref{capacityfig1} (\subref{capacityfignmu2}) it can be seen that the average capacity increases with decrease in the magnitude of $\eta$. Also, the average capacity increases with increase in the number of transmitters $N_T$, but the increase is diminished with larger $N_T$. For any further increase in capacity one has to increase either $N_R$ or the SNR, as shown in Fig. \ref{capacityfig1} (\subref{capacityfignmu1}).  It can be observed from Fig. \ref{capacityfig2} (\subref{capacityfigkmu1}), that capacity decreases with both $\kappa$ and $\mu$. 
\subsection{Outage probability of MIMO-MRC}
We  consider  a  wireless  link  equipped  with $N_T$ antenna  at the transmitter and $N_R$ antenna at the receiver. The received vector at the receiver can be modelled as
\begin{equation}
    \vec y = \vec H \vec w s+ \vec n,
\end{equation}
where $s$ is  the transmitted signal  of  the desired  user and $\vec n$ is the additive white Gaussian noise with power $\sigma_n^2$. $\vec w$ the weight vector at the transmitter with power $\Omega_D$ and $\vec H$ is the channel gain matrix. For MIMO-MRC, the maximum output SNR at the receiver is \cite[Eq. 27]{kang2003mrc}
\begin{equation}
   \gamma=\frac{\Omega_D}{\sigma_n^2}\lambda_{max},
\end{equation}
where $\lambda_{max}$ is the largest eigenvalue of $\vec{HH}^H$.
The outage probability is the CDF of the output SNR evaluated at $\gamma_{th}$.
\subsubsection{$\eta-\mu$}
Since $\vec{HH}^H \sim \mathcal{CW}_{N_R}(n, b \vec I_{N_R})$, the OP is directly given by \cite[Eq. 33]{kang2003mrc}
\begin{align}
    P_{out} &=\text{Pr}(\gamma \leq \gamma_{th})\\
    &= \frac{|\vec \Psi_c(\frac{\sigma_n^2 \gamma_{th}}{\Omega_D b} |}{\prod_{k=1}^{N_R}\Gamma(n-k+1)\Gamma(N_R-k+1)},
\end{align}
where $\{\vec \Psi_c(x) \}_{i,j}= \gamma(n-N_R+i+j-1,x)$ and $\gamma(.,.)$ is the incomplete gamma function. The above expression is valid for $N_R < N_T$. In case $N_T < N_R$, we approximate $\vec {H}^H \vec H$ instead of $\vec{HH}^H$ since both have the same non-zero eigenvalues.
\subsubsection{$\kappa-\mu$}
The outage expressions for $\kappa-\mu$ fading in MRC is derived in Appendix \ref{capacity}. For $N_T  \geq N_R$, by substituting $n_1=N_R$, $x= \frac{\gamma_{th} \sigma_n^2}{\Sigma_D}$, $w_1=2\sigma^2(1+\kappa) \mu-y$ and $w_2=  2 \sigma^2  (1+\kappa) \mu +(N_R-1)y$ which are the eigenvalues of $\vec \Sigma^{-1}$ in (\ref{kmu_outage}), we can get the outage probability approximation. In case $N_T < N_R$, we approximate $\vec {H}^H \vec H$ instead of $\vec{HH}^H$ since both have the same non-zero eigenvalues. For $N_T  \leq N_R$, we substitute $n_1=N_T$ in (\ref{kmu_outage}) to get the outage approximation. Here, $w_1=2\sigma^2(1+\kappa) \mu-y$ and $w_2=  2 \sigma^2  (1+\kappa) \mu +(N_T-1)y$ are the eigenvalues of $\vec \Sigma$ with multiplicity $N_T-1$ and $1$ respectively. 
\begin{figure}
\centering
\subcaptionbox{ Outage probability for $N_R=3$, $\eta=0.3$ and $\mu=3$ \label{outagefignmu1}}{\includegraphics[scale=0.4]{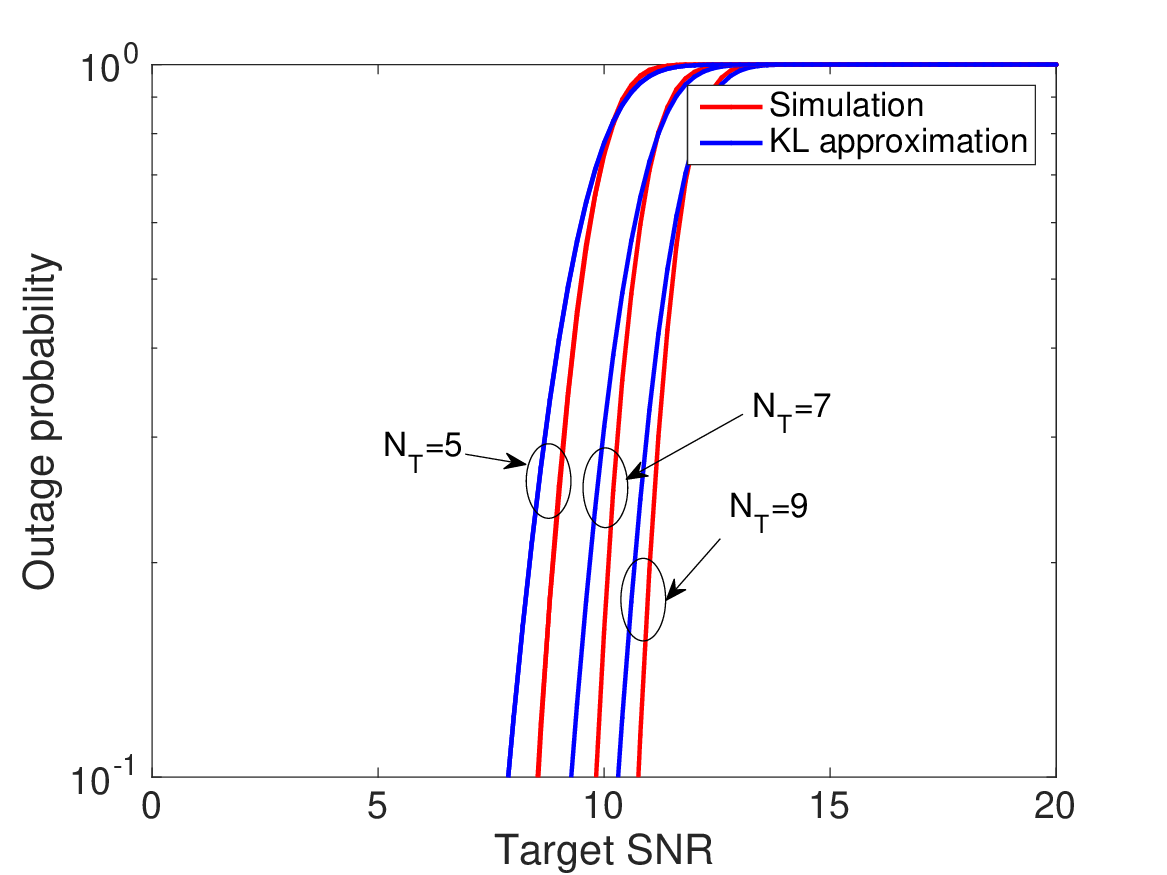}}
\hfill
\subcaptionbox{Outage probability for $N_R=2$, $\kappa=0.2$ and $\mu=2$ \label{outagefigkmu1}}{\includegraphics[scale=0.4]{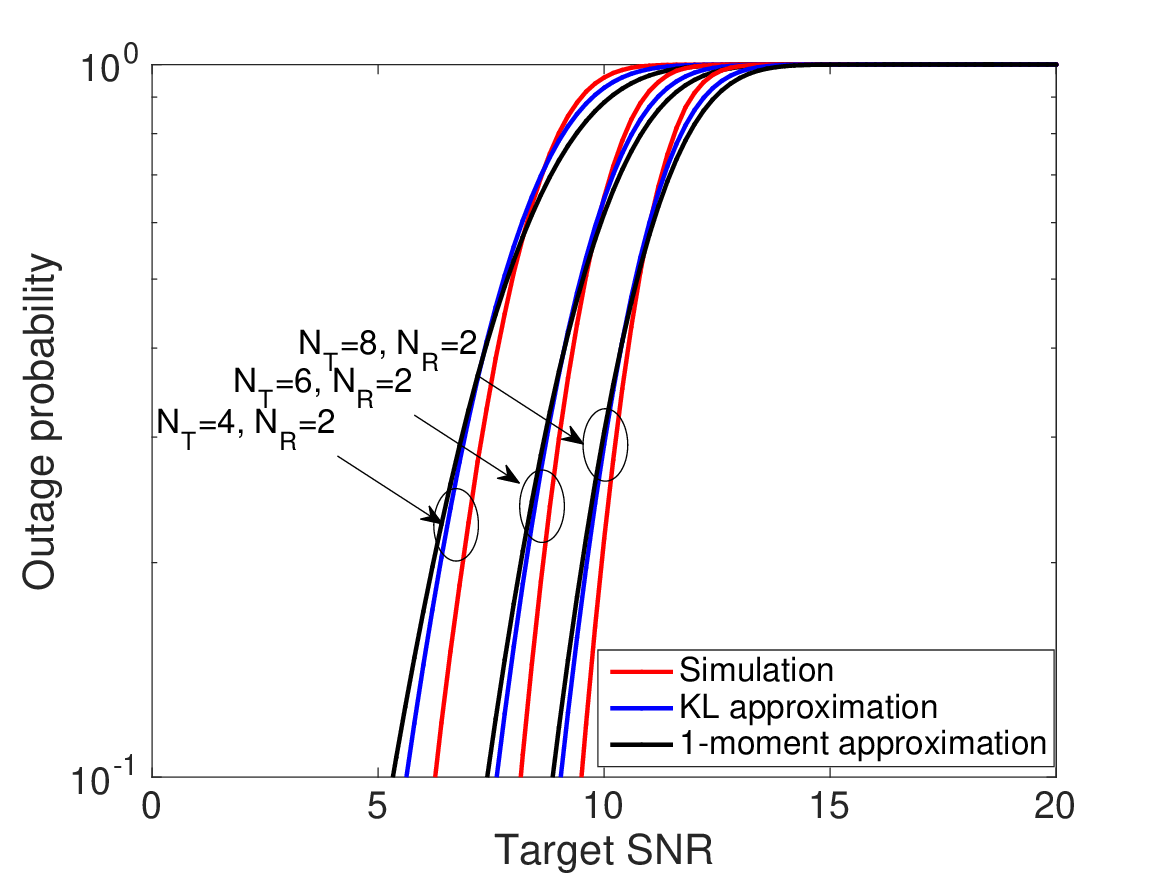}}
\caption{Outage probability of MRC for varying $N_T$}
\label{outagefig1}
\end{figure}
\par The derived capacity expressions are verified using Monte-Carlo simulations. A close match is found between the theoretical and simulation results for all the cases as can be seen from the Fig. \ref{outagefig1}. For both the cases it can be seen that the outage decreases with increase in the number of transmitting antennas $N_T$. The KL approximation, even though provides the Wishart distribution with the least KL divergence from the actual distribution, doesn't approximate the tail of the distribution perfectly. This is reflected by means of the small gap between the simulation and the approximation for outage in the order of $10^{-1}$. Nevertheless, the KL approximation performs better than the 1-moment approximation as seen from Fig. \ref{outagefig1}(\subref{outagefigkmu1}).

\subsection{Ergodic rate of ZF receiver in massive MIMO with low resolution ADCs}
Massive multi-input-multi-output (MIMO) system has been widely accepted as a key technology to meet the increasing demand for wireless throughput in both mobile and fixed scenarios and has been widely investigated in \cite{marzetta2015massive,bjornson2016massive, hanif2018antenna, gao2016antenna}. However, one drawback with massive MIMO systems is that, since a large number of antennas are required, there is a substantial increase in the hardware cost and power consumption. Using high-speed high-resolution analog-to-digital converter (ADC) for all the antennas increases the power consumption of massive MIMO systems severely and this is considered as the bottleneck to realize massive MIMO in practice. To solve the power consumption problem, typically low-resolution ADCs (e.g., 1-3 bits) are employed at the RF chains \cite{fan2015adc}.
\par Hence, it is imperative to study the performance of MIMO systems in conjunction with a quantizer. Works like \cite{singh2009adc, bai2013adc, orhan2015adc} have analyzed such quantized MIMO systems. Due to the complicated nature of the exact quantization error, the quantization is modeled as additive and independent noise. This additive quantization noise model (AQNM) is further used in \cite{fan2015adc} to study the impact of ADC resolution on the uplink rate for the case of Rayleigh fading channels. Some asymptotic results are also discussed in \cite{fan2015adc}. This was extended to the case of Rician fading in \cite{zhang2016adc}. In \cite{tan2016adc}, a mixed ADC architecture model is considered. In this architecture, a few antennas are equipped with costly full-resolution ADC and the rest with less expensive low-resolution ADC. This was further extended to the case of Rician fading channels in \cite{zhang2017adc} and Nakagami-m channels \cite{srinivasan2019analysis}. Very recently, the authors of \cite{ding2018mimo} have derived an approximate outage probability (OP) expression using the fact that if the squared coefficient of variation (SCV) of a random variable (RV) tends to zero, the RV approaches a deterministic value equal to its mean.

\par All the above works consider MRC at the receiver. There are a few works like \cite{qiao_adc, jacobsson_adc}, which consider ZF at the receiver . It is shown that the zero-forcing (ZF) receiver outperforms the MRC receiver in rate when the resolution of ADCs, the numbers of users (UEs) and BS antennas are fixed. In this work, we derive approximations for ergodic rate when the channels are $\kappa-\mu$ or $\eta-\mu$ faded and ZF is employed at the BS. The uplink of a multi-user MIMO (MU-MIMO) system formed by a BS with an array of $M$ antennas serving $N$ single-antenna user terminals is considered. All the user terminals are served in the same time-frequency resource.
The received $M \times 1$ vector $\vec y$ at BS can be expressed as
\begin{equation}\label{y}
\vec y = \vec G \vec x + \vec n,
\end{equation}
where $\vec G$ represents the $M \times N$ channel matrix between BS and users, $\vec x$ denotes unit-power $N \times 1$ symbols transmitted and $\vec n$ denotes the complex Gaussian noise i.e., $\vec n \sim \mathcal{CN}(0, \sigma^2\vec I)$.
The $ln$th element $g_{ln}$ of the matrix $\vec G$ is $\eta-\mu$ or $\kappa-\mu$ fading complex coefficient.
For AQNM, the output at the quantizer is given by \cite{fan2015adc}
\begin{equation}\label{yq}
\vec y_q = \alpha \vec y + \vec n_q =\alpha  \vec G \vec x + \alpha \vec n + \vec n_q,
\end{equation}
with $\alpha =1-\rho$, where $\rho$ is the inverse of the signal-to-quantization-noise ratio, and $\vec n_q$ is the additive Gaussian quantization noise vector that is uncorrelated with $\vec y$. The relation between the number of quantization bins denoted by $b$ and $\rho$ is given in \cite[Table I]{fan2015adc} for $b \leq 5$ and is approximated as $\rho = \frac{\pi \sqrt{3}}{2}2^{-2b}$ for $b > 5$.
For a fixed channel realization $\vec G$ and an identity input covariance matrix, $\vec{R_{n_q n_q}}$ is the covariance of $\vec n_q$ and is given by \cite[Eq. 5]{fan2015adc}
\begin{equation}\label{cov}
\vec{R_{n_q n_q}} =\alpha(1-\alpha) \text {diag}\left( \vec {GG}^H+ \sigma^2\vec I\right).
\end{equation}
We assume that a zero-forcing receiver (ZF) is implemented at the BS. At the output of the ZF receiver, the received signal is given by
\begin{equation}\label{r1}
\vec r = \left(\vec G \left(\vec {G}^H \vec G \right)^{-1}\right)^H \vec {y}_q.
\end{equation}
By substituting (\ref{yq}) in (\ref{r1}), we obtain
\begin{equation}\label{r2}
\vec r= \alpha \left(\vec G \left(\vec {G}^H \vec G \right)^{-1}\right)^H\vec G \vec x+ \alpha \left(\vec G \left(\vec {G}^H \vec G \right)^{-1}\right)^H\vec n + \left(\vec G \left(\vec {G}^H \vec G \right)^{-1}\right)^H \vec n_q.
\end{equation}
The $n^{th}$ element of $\vec r$ is given by
\begin{equation}\label{rn}
r_n = \alpha x_n  + \alpha \vec {\tilde g_n}^H \vec n +\vec {\tilde g_n}^H \vec n_q,
\end{equation}
where $\vec {\tilde g_n}=\left[\vec G \left(\vec {G}^H \vec G \right)^{-1}\right]_{:, n}$ is the filter for $n$th user. Hence, the SIQNR of the $n$th user is given by
\begin{align}\label{siqnr}
\gamma_n&= \frac{X_n}{Y_n}= \frac{ \alpha^2 }{ \alpha^2 ||\vec {\tilde g_n}||^2 + \alpha(1-\alpha)Z_n },
\end{align}
where $Z_n=\vec {\tilde g_n}^H diag( \vec {GG}^H+ \sigma^2\vec I)\vec {\tilde g_n}$. 
Now using the popular approximation from \cite{fan2015adc, zhang2016adc, zhang2017adc},
the approximate ergodic rate is given by
\begin{align}
    \text{Rate} &= E[\text{log}_2(1+\gamma_n)]\\
    &=  E\left[\text{log}_2\left(1+\frac{ \alpha^2 }{ \alpha^2 ||\vec {\tilde g_n}||^2 + \alpha(1-\alpha)Z_n }\right)\right]\\
    & \approx \text{log}_2\left(1+\frac{ \alpha^2 }{ E[\alpha^2 ||\vec {\tilde g_n}||^2 + \alpha(1-\alpha)Z_n] }\right).
\end{align}
Note that, since $\vec {\tilde g_n}=\left[\vec G \left(\vec {G}^H \vec G \right)^{-1}\right]_{:, n}$, $||\vec {\tilde g}_n||^2=\left[ \left(\vec {G}^H \vec G \right)^{-1}\right]_{n, n}$. For large $M$, from \cite{liuarxiv}, the following approximation identity holds $\vec {\tilde g}_n^H \vec R_{n_qn_q} \vec {\tilde g}_n \approx \vec {\tilde g}_n^H E[\vec R_{n_qn_q}] \vec {\tilde g}_n$. Therefore,
\begin{align}
\nonumber
    E[Z_n] &\approx E\left[\vec {\tilde g}_n^H E[\vec R_{n_qn_q}] \vec {\tilde g}_n\right]\\
    \nonumber
     &= E\left[\vec {\tilde g}_n^H E[diag( \vec {GG}^H+ \sigma^2\vec I)] \vec {\tilde g}_n\right]\\
     &= E\left[||\vec {\tilde g}_n||^2\right] \left[E(\vec{GG}^H)_{1,1}+  \sigma^2\right],
\end{align} 
where $E(\vec{GG}^H)_{1,1}$ is the expectation of the diagonal element of $\kappa-\mu$ or $\eta-\mu$ faded channel and is already determined in Section III.B. Therefore,
\begin{align}
    \text{Rate} & \approx \text{log}_2\left(1+\frac{ \alpha^2 }{ E[\alpha^2 ||\vec {\tilde g_n}||^2] + \alpha(1-\alpha)E\left[||\vec {\tilde g}_n||^2\right] \left[E(\vec{GG}^H)_{1,1}+  \sigma^2\right] }\right). \label{rate}
\end{align}
For $\eta-\mu$ channel, $E(\vec{GG}^H)_{1,1}= K (\Omega_X + \Omega_Y)$ and for $\kappa-\mu$ channel, $E(\vec{GG}^H)_{1,1}= K 2\sigma^2  (1+\kappa)\mu$. Once $E[||\vec {\tilde g_n}||^2]$ is determined, we can determine the expressions for rate. But it is intractable to exactly determine $E[||\vec {\tilde g_n}||^2]$ for $\kappa-\mu$ or $\eta-\mu$ faded elements of $G$. With our Wishart approximation, one can use Wishart properties to determine $E[ ||\vec {\tilde g_n}||^2]$. $\vec{G}^H\vec G$ is approximated by a Wishart matrix $\mathcal{CW}_{K}(m, \vec \Sigma)$, where $\vec \Sigma$ and $m$ are obtained by (\ref{sigma1}) and (\ref{nequation}). Therefore, using the Wishart identity $E\left[\left[ \left(\vec {G}^H \vec G \right)^{-1}\right]_{n, n}\right] \approx \frac{\left[\Sigma^{-1}\right]_{n, n}}{m-K}$, one can substitute $E[||\vec {\tilde g_n}||^2]$ in (\ref{rate}) to determine the approximate rate expressions.
\begin{figure}
\centering
\subcaptionbox{Ergodic rate for $K=5$, $\eta=0.1$ and $\mu=1$ \label{ratefignmu1}}{\includegraphics[scale=0.4]{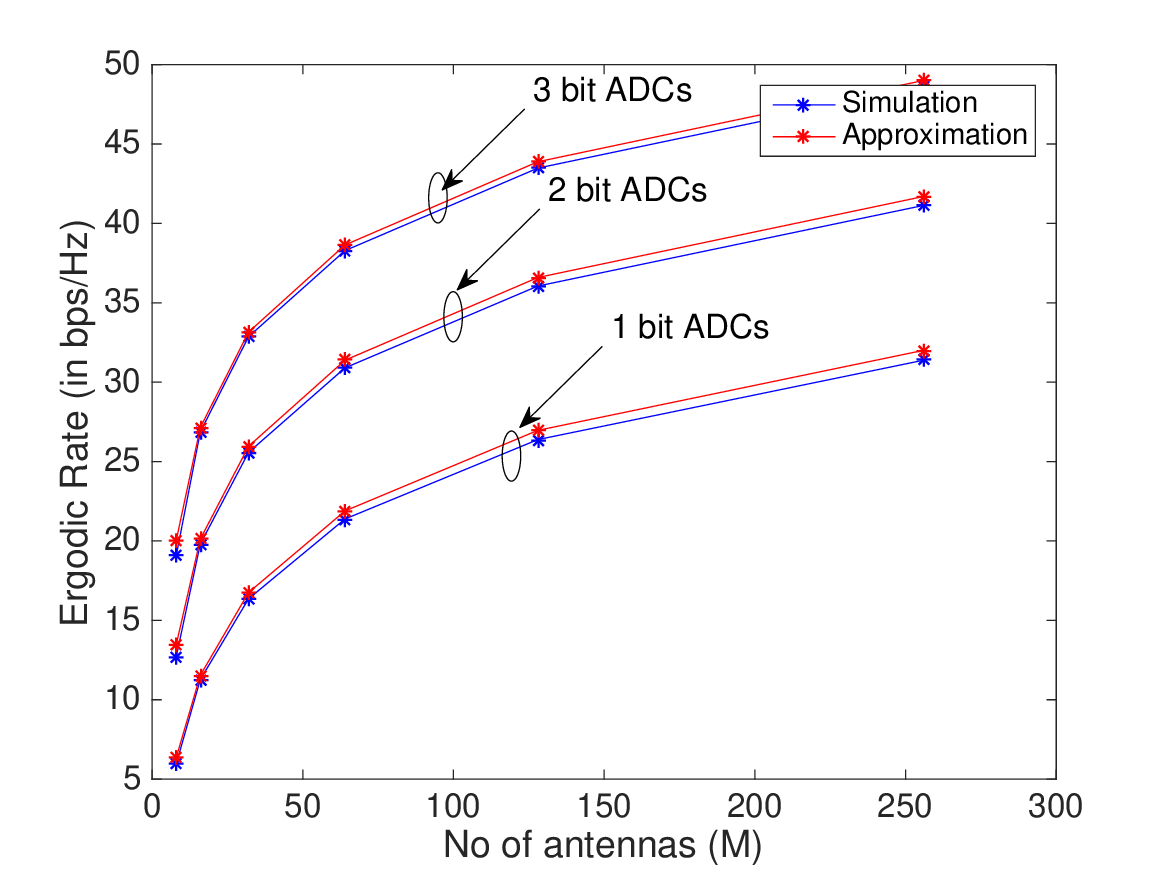}}
\hfill
\subcaptionbox{Ergodic rate for $K=5$, $\kappa=0.2$ and $\mu=2$ \label{ratefigkmu1}}{\includegraphics[scale=0.4]{rate_vs_M_K5_mu1_eta1.eps}}
\caption{Ergodic rate expressions for varying $K$ and $M$}
\label{ratefig1}
\end{figure}
\par The derived ergodic rate expressions are verified using Monte-Carlo simulations. A close match is found between the theoretical and simulation results for all the cases as can be seen from the Fig. \ref{ratefig1}. For both the cases it can be seen that the ergodic rate increases with increase in the number of transmitting antennas $M$. This is because of the increase in diversity from the users to the BS. Also, the ergodic rate increases with better resolution of ADCs.

\section{Conclusions and Future work}
Approximate random matrix models have been derived for $\vec{HH}^H$ when the elements of $\vec H$ are i.i.d $\kappa-\mu$ or $\eta-\mu$ random variables. The approximation is terms of the complex Wishart matrix, which has the least KL-divergence from the original matrix distribution. Further, for $\kappa-\mu$ distribution, we also derive a simple approximation that matches the first moment but fixes the degree of freedom. The utility of our result is shown by a) deriving approximate capacity expressions for $\kappa-\mu$ or $\eta-\mu$ MIMO models b) deriving approximate outage expressions of MIMO-MRC for $\kappa-\mu$ or $\eta-\mu$ channels c) deriving ergodic rate expressions for zero-forcing (ZF) receiver in an uplink single cell massive MIMO scenario with low resolution analog-to-digital converters (ADCs) in the antennas. For these applications, extensive Monte-Carlo simulations have been performed and an excellent match with the approximate expressions has been observed. Given the simplicity of the approximation, it can also be extended to other types of generalized fading matrix models.

\appendices
\section{Approximate mean of complex $\kappa-\mu$ random variables}\label{off_kmu}
The expectations to be approximated are,
\begin{align}\label{expx}
  E[x_{ik}]= \int_{- \infty}^{\infty} x \frac{|x|^{\mu/2}}{2\vec \sigma^2 |p|^{\mu/2-1}}exp(-\frac{(x-p)^2}{2 \vec \sigma^2})sech(\frac{px}{\vec \sigma^2})I_{\frac{\mu}{2}-1}( \frac{  |p x|}{\vec \sigma^2} )dx
\end{align}
\begin{align}\label{expy}
  E[y_{ik}]= \int_{- \infty}^{\infty} y \frac{|y|^{\mu/2}}{2\vec \sigma^2 |q|^{\mu/2-1}}exp(-\frac{(y-q)^2}{2 \vec \sigma^2})sech(\frac{qy}{\vec \sigma^2})I_{\frac{\mu}{2}-1}( \frac{  |qy|}{\vec \sigma^2} )dy.
\end{align}
The expectation $  E[x_{ik}^r]$, for some $r>0$, using the trigonometric identity $tanh(z)=1-e^{-z}sech(z)$, is given as,
\begin{align*}
  E[x_{ik}^r]=2 \int_{0}^{\infty}  \frac{x^{\mu/2+r}}{2\vec \sigma^2 |p|^{\mu/2-1}}exp(-\frac{x^2}{2 \vec \sigma^2})exp(-\frac{p^2}{2 \vec \sigma^2})tanh(\frac{px}{\vec \sigma^2})I_{\frac{\mu}{2}-1}( \frac{  p x}{\vec \sigma^2} )dx.
\end{align*}
Since, the above integral cannot be solved to obtain a solution in closed form, we can approximate $tanh(\frac{px}{\vec \sigma^2})$ by $erf(\frac{\sqrt{\pi}}{2}\frac{px}{\vec \sigma^2})$ to obtain \cite{reyes},
\begin{align*}
  E[x_{ik}^r] \approx 2 \int_{0}^{\infty}  \frac{x^{\mu/2+r}}{2\vec \sigma^2 |p|^{\mu/2-1}}exp(-\frac{x^2}{2 \vec \sigma^2})exp(-\frac{p^2}{2 \vec \sigma^2})erf(\frac{\sqrt{\pi}}{2} \frac{px}{\vec \sigma^2})I_{\frac{\mu}{2}-1}( \frac{  p x}{\vec \sigma^2} )dx.
\end{align*}
Using the identity $I_v(z)= \frac{1}{\Gamma(v+1)}(\frac{z}{2})^v \,_0F_1(v+1, \frac{z^2}{4})$ from \cite{abr},
\begin{align*}
  E[x_{ik}^r]\approx 2 \int_{0}^{\infty}  \frac{x^{\mu/2+r}}{2\vec \sigma^2 |p|^{\mu/2-1}}exp(-\frac{x^2}{2 \vec \sigma^2})exp(-\frac{p^2}{2 \vec \sigma^2})erf(\frac{\sqrt{\pi}}{2} \frac{px}{\vec \sigma^2})\frac{1}{\Gamma(\mu/2)} (\frac{px}{2 \sigma^2})^{\mu/2-1}\,_0F_1(\frac{\mu}{2},\frac{  p^2 x^2}{4 \sigma^4} )dx.
\end{align*}
Now, expanding the hypergeometric series and interchanging the integration and summation, we obtain,
\begin{align*}
  E[x_{ik}^4]&\approx 2 \int_{0}^{\infty}  \frac{x^{\mu/2+r}}{2\vec \sigma^2 |p|^{\mu/2-1}}exp(-\frac{x^2}{2 \vec \sigma^2})exp(-\frac{p^2}{2 \vec \sigma^2})erf(\frac{\sqrt{\pi}}{2} \frac{px}{\vec \sigma^2})\frac{1}{\Gamma(\mu/2)} (\frac{px}{2 \sigma^2})^{\mu/2-1}\sum_{n=0}^{\infty}\frac{1}{(\frac{\mu}{2})_n n!}(\frac{  p^2 x^2}{4 \sigma^4} )^ndx\\
&= 2 \sum_{n=0}^{\infty}\frac{p^{2n}}{(\frac{\mu}{2})_n n!}e^{-\frac{p^2}{2  \sigma^2}} \frac{1}{\Gamma(\mu/2)} \frac{1}{(2 \sigma^2)^{\mu/2+ 2n} }\int_{0}^{\infty}x^{\mu +2n+r-1}exp(-\frac{x^2}{2  \sigma^2})erf(\frac{\sqrt{\pi}}{2} \frac{px}{ \sigma^2})dx
\end{align*}
Applying the integration identity  $\int_{0}^{\infty}erf(ax) e^{-b^2 x^2} x^p dx= \frac{a}{\sqrt{\pi}}b^{-p-2} \Gamma( \frac{p}{2}+1) \,_2F_1(\frac{1}{2}, \frac{p}{2}+1, \frac{3}{2}, -\frac{a^2}{b^2})$ for $b^2 >0$ and $ p >-2$ from \cite{ng_table}, we obtain,
\begin{align*}
  E[x_{ik}^r]&=  2pe^{-\frac{p^2}{2  \sigma^2}}\sum_{n=0}^{\infty}  p^{2n}(\frac{1}{2 \sigma^2})^{n-r/2+1/2} \frac{1}{ (\frac{\mu}{2})_n n! \Gamma(\mu/2)} \Gamma(\mu/2+n+r/2-1/2+1)\\
  & \times \,_2F_1(\frac{1}{2}, \mu/2+n+r/2-1/2+1, \frac{3}{2}, -\frac{2 p^2}{\sigma^2} \frac{\pi}{4}).
\end{align*}
Using the transformation $ \,_2F_1(a,b,c,z)= (1-z)^{-b} \,_2F_1(c-a, b,c, \frac{z}{z-1})$ for the Gauss Hypergeometric function from \cite{for}, we obtain,
\begin{align}
\nonumber
  E[x_{ik}^r] &\approx  2pe^{-\frac{p^2}{2  \sigma^2}}\sum_{n=0}^{\infty} p^{2n}(\frac{1}{2 \sigma^2})^{n-r/2+1/2}\frac{\Gamma(\mu/2+n+r/2-1/2+1)}{ (\frac{\mu}{2})_n n! \Gamma(\mu/2)}  (1+\frac{2 p^2}{\sigma^2} \frac{\pi}{4})^{ -\mu/2-n-1-r/2+1/2} \\
  \nonumber
  & \,_2F_1(1, \mu/2+n+r/2-1/2+ 1, \frac{3}{2}, \frac{2 p^2 \pi}{2 p^2 \pi+4 \sigma^2})\\
  \nonumber
&= 2pe^{-\frac{p^2}{2  \sigma^2}}(\frac{4 \sigma^2}{4 \sigma^2+2p^2 \pi})^{\mu/2 +1} 2^{r-1}\left(\frac{1}{4\sigma^2+2p^2\pi}\right)^{r/2-1/2}\\
& \sum_{n=0}^{\infty}  ( \frac{2 p^2}{ 4 \sigma^2+ 2p^2 \pi})^n \frac{\Gamma(\mu/2+n+r/2-1/2+1)}{ (\frac{\mu}{2})_n n! \Gamma(\mu/2)} \,_2F_1(1, \mu/2+n+r/2-1/2+1, \frac{3}{2}, \frac{2 p^2 \pi}{2 p^2 \pi+4 \sigma^2}). \label{xpowerr}
\end{align}
Similarly,
\begin{align}
  \nonumber
E[y_{ik}^r]&= 2qe^{-\frac{q^2}{2  \sigma^2}}(\frac{4 \sigma^2}{4 \sigma^2+2q^2 \pi})^{\mu/2 +1} 2^{r-1}\left(\frac{1}{4\sigma^2+2q^2\pi}\right)^{r/2-1/2}\\
& \sum_{n=0}^{\infty}  ( \frac{2 q^2}{ 4 \sigma^2+ 2q^2 \pi})^n \frac{\Gamma(\mu/2+n+r/2-1/2+1)}{ (\frac{\mu}{2})_n n! \Gamma(\mu/2)} \,_2F_1(1, \mu/2+n+r/2-1/2+1, \frac{3}{2}, \frac{2 q^2 \pi}{2 q^2 \pi+4 \sigma^2}).\label{ypowerr}
\end{align}
We can use the above expression to compute $E[x_{ik}^r]$ and $E[y_{ik}^r]$ or the expression can be further simplified further in terms of the Appell function. For example, for $r=1$, expanding the $\,_2F_1$ as series
\begin{align*}
  E[x_{ik}] &=  2pe^{-\frac{p^2}{2  \sigma^2}}(\frac{4 \sigma^2}{4 \sigma^2+2p^2 \pi})^{\mu/2 +1}\frac{\Gamma(\mu/2+1)}{\Gamma(\mu/2)}\sum_{n=0}^{\infty} \sum_{k=0}^{\infty}    \frac{ (\mu/2+1)_{n+k} (1)_k }{ (\frac{3}{2})_k (\frac{\mu}{2})_n  } \frac{( \frac{2 p^2}{ 4 \sigma^2+ 2p^2 \pi})^n}{n!} \frac{ (\frac{2 p^2 \pi}{2 p^2 \pi+4 \sigma^2})^k}{k!}.
\end{align*}
Rewriting the above using  confluent Appell function $\Psi_1$ \cite{brychkov},
\begin{align}\label{xapprox}
\nonumber
  E[x_{ik}] &\approx  2pe^{-\frac{p^2}{2  \sigma^2}}(\frac{4 \sigma^2}{4 \sigma^2+2p^2 \pi})^{\mu/2 +1}\frac{\Gamma(\mu/2+1)}{\Gamma(\mu/2)}\\
& \qquad \qquad \Psi_1( \mu/2+1,1,3/2, \mu/2,\frac{2 p^2 \pi}{2 p^2 \pi+4 \sigma^2},  \frac{2 p^2}{ 4 \sigma^2+ 2p^2 \pi}).
\end{align}
Similarly,
\begin{align}\label{yapprox}
\nonumber
  E[y_{ik}] &\approx 2qe^{-\frac{q^2}{2  \sigma^2}}(\frac{4 \sigma^2}{4 \sigma^2+2q^2 \pi})^{\mu/2 +1}\frac{\Gamma(\mu/2+1)}{\Gamma(\mu/2)}\\
& \qquad \qquad \Psi_1( \mu/2+1,1,3/2, \mu/2,\frac{2 q^2 \pi}{2 q^2 \pi+4 \sigma^2},\frac{2 q^2}{ 4 \sigma^2+ 2q^2 \pi}).
\end{align}
We have compared (\ref{xapprox}) and (\ref{yapprox}) with numerical evaluation of the expectation integrals and also empirical average of simulated $\kappa-\mu$ variables for a wide range of parameters. In all cases, an excellent match has been observed.

\section{Determining $tr \left[E_p\left( E_p[\vec X]^{-1}\vec X E_p[\vec X]^{-1}\vec X\right)\right]$ for $\vec X$}\label{2by2}
Assume $\vec H$ is of dimension $2 \times N_T$ and $ij$th element ($i=1,2$ and $j=1,..., N_T$) of $\vec H$ is $x_{ij}+y_{ij}$ and a complex $\kappa-\mu$ RV.  $\vec X= \vec {HH}^H$ is of dimension $2 \times 2$. Denote $\vec X$ by $\begin{bmatrix}
a & b\\ 
\bar b & c
\end{bmatrix}$. Now to evaluate $tr \left(E_p\left( E_p[\vec X]^{-1}\vec X E_p[\vec X]^{-1}\vec X\right)\right)$, we substitute $E_p[X]=Z= (z_{diag}-z_{off})\vec I_{n_1} + z_{off}\vec1\vec1^T$, where $z_{diag}$ and $z_{off}$ are given by (\ref{zdiag}) and (\ref{zij}) respectively. Applying this substitution and using Sherman-Morrison formula for $Z^{-1}$, we obtain
\begin{align}
    Z^{-1}X &= (z_{diag}-z_{off})^{-1} X - \frac{z_{off}(z_{diag}-z_{off})^{-2}  \vec 1 \vec 1^T X }{1+ n_1 z_{off} (z_{diag}-z_{off})^{-1} }
\end{align}
Therefore,
\begin{align}
\nonumber
    (Z^{-1}X)^2 &= (z_{diag}-z_{off})^{-2} X^2 + \frac{z_{off}^2(z_{diag}-z_{off})^{-4}  (\vec 1 \vec 1^T X)^2 }{(1+ n_1 z_{off} (z_{diag}-z_{off})^{-1})^2 }\\
    &-   \frac{z_{off}(z_{diag}-z_{off})^{-3}  X \vec 1 \vec 1^T X }{1+ n_1 z_{off} (z_{diag}-z_{off})^{-1} } -   \frac{z_{off}(z_{diag}-z_{off})^{-3}  \vec 1 \vec 1^T X X}{1+ n_1 z_{off} (z_{diag}-z_{off})^{-1} }. 
\end{align}
Hence, to determine $tr\left[ E_p\left( (\vec Z^{-1}\vec X)^2\right)\right]$, we need to evaluate $E_p[tr(\vec X^2)]$, $E_p[tr((\vec{11}^T \vec X)^2)]$ and $E_p[tr(\vec{11}^T \vec X^2)]$. Substituting $\vec X= \begin{bmatrix}
a & b\\ 
\bar b & c
\end{bmatrix}$, we obtain:
\begin{align}
    E_p[tr(\vec X^2)] &= E_p\left[ a^2+ c^2+ 2b\bar b \right],
\end{align}
\begin{align}
\nonumber
    E_p[tr((\vec{11}^T \vec X)^2)] &= E_p\left[ (a+ \bar b)^2+ 2(a+\bar b)(b+c) + (b+c)^2 \right]\\
    &= E_p\left[a^2+ \bar b^2+ 2a\bar b + 2ab+2\bar b b+2ac + 2\bar b c + b^2+c^2+2bc \right]
\end{align}
and
\begin{align}
    E_p[tr(\vec{11}^T \vec X^2)] &= E_p\left [ a^2+ 2b\bar b +c^2 + ab +a \bar b +bc + \bar b c\right].
\end{align}
The terms in the above equations can be trivially evaluated by the substituting the terms of $\vec X$ in terms of the elements of $\vec H$ and simplifying. For a $\vec H$ of dimension $2 \times N_T$, we obtain
\begin{equation}
    E_p[a^2]= E_p[c^2]=N_T(E_p[x_{ik}^4]+ E_p[y_{ik}^4])+  (N_T-1)(E_p[x_{ik}^2]+E_p[y_{ik}^2])^2 + 2(E_p[x_{ik}^2]E_p[y_{ik}^2])
\end{equation}
\begin{equation}
   E_p[b \bar b]= N_T((E_p[x_{ik}^2]+E_p[y_{ik}^2])^2  +(N_T-1)(E_p[x_{ik}]^2+E_p[y_{ik}]^2)^2)
\end{equation}
\begin{align}
\nonumber
    E_p[ab] = E_p[cb]&= N_T(E_p[x_{ik}^3]E_p[x_{ik}] + N_TE_p[x_{ik}^2]E_p[y_{ik}]^2\\ 
    \nonumber
    & +(N_T-1)E_p[x_{ik}^2]E_p[x_{ik}]^2-jE_p[x_{ik}^3]E_p[y_{ik}]+jE_p[x_{ik}^2]E_p[x_{ik}]E_p[y_{ik}]\\
    \nonumber
    &+ E_p[y_{ik}^3]E_p[y_{ik}] + N_TE_p[y_{ik}^2]E_p[x_{ik}]^2\\ &+(N_T-1)E_p[y_{ik}^2]E_p[y_{ik}]^2+jE_p[y_{ik}^3]E_p[x_{ik}]-jE_p[y_{ik}^2]E_p[y_{ik}]E_p[x_{ik}]
\end{align}
where $E_p[x_{ik}]$, $E_p[x_{ik}^2]$, $E_p[x_{ik}^3]$ and $E_p[x_{ik}^4]$ can be obtained using (\ref{xpowerr}) and $E_p[y_{ik}]$, $E_p[y_{ik}^2]$, $E_p[y_{ik}^3]$ and $E_p[y_{ik}^4]$ can be obtained using (\ref{ypowerr}).

\section{Capacity and outage for $\kappa-\mu$}\label{capacity}
\subsection{Capacity}
We have to determine an approximation for $C=E_{\vec \Lambda}[\sum_{i=1}^{n_1} log_2(1+ \frac{\rho}{N_T} \lambda_i)]$, where $\lambda_k$ for $k=1,.., n_1$ are eigenvalues of a $n_1 \times n_1$ random matrix $\vec R= \vec{HH}^H$, where $\vec H$ have i.i.d. $\kappa-\mu$ or $\eta-\mu$ elements.
 We approximate the matrix $\vec R$ by a  $n_1 \times n_1$ central Wishart matrix $\vec W \sim \mathcal{CW}_{n_1}(n_2, \vec \Sigma)$, such that $n_1 \leq n_2$ and $\vec \Sigma$ as in (\ref{kmusigma}). The eigenvalue distribution of the unordered eigenvalues of $\vec W$ is given by,
\begin{align}\label{eig}
f(\vec \Lambda)= (-1)^{\frac{1}{2} n_1(n_1-1)} \frac{1}{n_1!} \frac{ det( \{ e^{-\lambda_i w_j} \} )}{| \vec \Sigma|^{n_2}} \frac{ \Delta( \vec \Lambda)}{\Delta( \vec \Sigma^{-1})} \prod_{j=1}^{n_1} \frac{ \lambda_j^{n_2-n_1}}{ (n_2-j)!}
\end{align}
where $ w_1 > w_2 > .... > w_{n_1}$ are the eigenvalues of $\vec \Sigma^{-1}$ and $\lambda_1,..., \lambda_{n_1}$ are the eigenvalues of $\vec W$. But if some eigenvalues of $\vec \Sigma^{-1}$ are not distinct, then the above distribution cannot be used because $det( \{ e^{-\lambda_i w_j} \} )= \Delta( \vec \Sigma^{-1})=0$ leading to an indeterminate form. Hence, we apply the following theorem from \cite{randombook}, to modify the distribution and account for non-distinct eigenvalues.
\begin{theorem}
Let $f_1,..., f_N$ be a family of infinitely differentiable functions and let $x_1,..., x_N \in \mathcal{R}$. Denote
$R(x_1,.., x_N) \triangleq \frac{det \big( \{f_i(x_j) \} \big)}{ \prod_{i<j} (x_j-x_i)}$.
Then, for $N_1,..., N_p$ such that $N_1+ ... +N_p= N$ and for $y_1,..., y_p \in R$ distinct, 
\begin{multline*}
\lim_{\substack{x_1,..., x_{N_1} \to y_1 \\ ....\\ x_{N-N_p+1,.., x_N} \to y_p}} R(x_1,..., x_N)\\
= \frac{det \big[ f_i(y_1), f_i'(y_1),..., f_i^{(N_1-1)}(y_1),..., f_i(y_p), f_i'(y_p),...., f_i^{(N_p-1)}(y_p) \big]}{ \prod_{1 \leq i < j \leq p}(y_j-y_i)^{N_i N_j} \prod_{l=1}^{p} \prod_{j=1}^{N_l-1} j!}.
\end{multline*}
\end{theorem}
In our case, $\vec \Sigma^{-1} $ has two eigenvalues $w_1$ and $w_2$ with multiplicity $n_1-1$ and $1$ respectively.
Hence, applying the above theorem to (\ref{eig}), we obtain, the eigenvalue distribution as,
\begin{align}\label{twoeig}
f(\vec \Lambda)= \frac{(-1)^{\frac{1}{2} n_1(n_1-1)} }{n_1!} \frac{ det( \{ e^{-\lambda_i w_1} \: (-\lambda_i)e^{-\lambda_i w_1} \: ... \: (-\lambda_i)^{n_1-2} e^{-\lambda_i w_1} \: e^{-\lambda_i w_2} \} )}{(w_2-w_1)^{n_1-1} \prod_{j=1}^{n_1-2} j!} \frac{ \Delta( \vec \Lambda)}{| \vec \Sigma|^{n_2}} \prod_{j=1}^{n_1} \frac{ \lambda_j^{n_2-n_1}}{ (n_2-j)!}.
\end{align}
Hence 
\begin{align*}
C  &\approx (-1)^{\frac{1}{2} n_1(n_1-1)} \frac{1}{n_1! \prod_{j=1}^{n_1} (n_2-j)! } \frac{1}{| \vec \Sigma|^{n_2} (w_2-w_1)^{n_1-1} \prod_{j=1}^{n_1-2} j!}  \\
 & \quad  \times \int_{0}^{\infty}[\sum_{i=1}^{n_1} log_2(1+ \frac{\rho}{N_T} \lambda_i)]  \lambda_k^{n_2-n_1} \Delta( \vec \Lambda) det( \{ e^{-\lambda_i w_1} \: (-\lambda_i)e^{-\lambda_i w_1} \: ... \: (-\lambda_i)^{n_1-2} e^{-\lambda_i w_1} \: e^{-\lambda_i w_2} \} )d \vec \Lambda.
\end{align*}
From Theorem 3 in Appendix of \cite{chiani_capacity}, it can be observed that,
for two arbitrary $ {n_1} \times {n_1}$ matrices $ \vec \Phi(\vec y) $ and $ \vec \Psi(\vec y)$  with $ij^{th}$  elements  $  \phi_i( y_j) $ and $  \Psi_i( y_j) $, and two arbitrary functions $ \xi(.)$ and $\xi'(.)$, where $\vec{y} = [y_1 \, y_2 \, ... \, y_{n_1}]^T $, the following identity holds:
\begin{align}\label{theorem3}
\nonumber
\textstyle \int \cdots  \int _{d \geq y_i \geq c} |\vec \Phi(\vec y)||\vec \Psi(\vec y)|\prod_{n=1}^{N} \xi(y_n) \sum_{k=1}^{N} \xi'(y_k) dy_1..dy_{n_1}  \textstyle \\  
\qquad ={N}!   \sum_{k=1}^{N} det \left( \left\{ \int_c^d \phi_i( y)   \Psi_j( y) \xi(y) U_{k,j}(\xi'(y)) dy\right\}_{1 \leq i,j \leq n_1} \right),
\end{align}
where,
$U_{j,k}(x)= x, \quad if \: k = j$ and $U_{j,k}(x)=1, \quad  if \: k \neq j.$
Applying the above identity, we obtain,
\begin{align}\label{kmu_capacity}
C  &\approx (-1)^{\frac{1}{2} n_1(n_1-1)} \frac{1}{ \prod_{j=1}^{n_1} (n_2-j)! } \frac{1}{| \vec \Sigma|^{n_2} (w_2-w_1)^{n_1-1} \prod_{j=1}^{n_1-2} j!}\sum_{k=1}^{n_1}det(N^k].
\end{align}
where 
\begin{align*}
\vec N^k_{i,j}(n_1, n_2) =\begin{cases}
 \int_{0}^{\infty}  \lambda^{n_2-n_1} \lambda^{i-1} (-\lambda)^{j-1}  e^{-\lambda w_1} d\lambda;  \quad   1  \leq i \leq n_1 , 1 \leq j  \leq n_1-1, j \neq k,\\
 \int_{0}^{\infty} log_2(1+ \frac{\rho}{N_T} \lambda) \lambda^{n_2-n_1} \lambda^{i-1} (-\lambda)^{j-1}  e^{-\lambda w_1} d\lambda;  \quad   1  \leq i \leq n_1 , 1 \leq j  \leq n_1-1, j =k,\\
  \int_{0}^{\infty} \lambda^{n_2-n_1} \lambda^{i-1}e^{-\lambda w_2} d\lambda;  \quad   1  \leq i \leq n_1 , j= n_1, j \neq k,\\
  \int_{0}^{\infty}log_2(1+ \frac{\rho}{N_T} \lambda) \lambda^{n_2-n_1} \lambda^{i-1}e^{-\lambda w_2} d\lambda;  \quad   1  \leq i \leq n_1 , j= n_1, j =k.
 \end{cases}
\end{align*}
First writing the logarithm in terms of Meijer-G function using the identity $ln(1+x)= \MeijerG*{1}{2}{2}{2}{1,1}{1,0}{x}$ \cite{for2} and solving the integrals using identities $\int_{0}^{\infty} x^{v-1}e^{-\mu x}dx= \Gamma(v) \mu^{-v}$ and $\int_{0}^{\infty}x^{-\rho}e^{-\beta x} $ $ \MeijerG*{m}{n}{p}{q}{a_1, \hdots, a_p}{b_1 ,\hdots ,b_q}{ \alpha x }dx= \beta^{\rho-1} \MeijerG*{m}{n+1}{p+1}{q}{\rho, a_1, \hdots, a_p}{b_1 ,\hdots ,b_q}{ \frac{\alpha}{\beta}  }$ from \cite{int}, we obtain,
\begin{align}\label{Nk}
\vec N^k_{i,j}(n_1, n_2) =\begin{cases}
 (-1)^{j-1}\Gamma(n_2-n_1+i+j-1)w_1^{-n_2+n_1-i-j+1};  \quad   1  \leq i \leq n_1 , 1 \leq j  \leq n_1-1, j \neq k,\\
 (-1)^{j-1} \frac{1}{ln2} \MeijerG*{1}{3}{3}{2}{1-n_2+n_1-i-j+1,1,1}{1,0}{ \frac{\rho}{N_T w_1} } w_1^{-n_2+n_1-i-j+1};  \\
 \qquad \qquad  1  \leq i \leq n_1 , 1 \leq j  \leq n_1-1, j =k,\\
\Gamma(n_2-n_1+i)w_2^{-n_2+n_1-i} ;  \quad   1  \leq i \leq n_1 , j= n_1, j \neq k,\\
\frac{1}{ln2} \MeijerG*{1}{3}{3}{2}{1-n_2+n_1-i,1,1}{1,0}{ \frac{\rho}{N_T w_2} } w_2^{-n_2+n_1-i};  \quad   1  \leq i \leq n_1 , j= n_1, j=k.
 \end{cases}
\end{align}
\subsection{Outage probability of MRC}
For the outage probability of MRC, from (\ref{twoeig}),
\begin{align*}
P(\lambda_{max}(W) <x)  &\approx (-1)^{\frac{1}{2} n_1(n_1-1)} \frac{1}{n_1! \prod_{j=1}^{n_1} (n_2-j)! } \frac{1}{| \vec \Sigma|^{n_2} (w_2-w_1)^{n_1-1} \prod_{j=1}^{n_1-2} j!}  \\
 & \quad  \times \int_{0}^{x}\lambda_k^{n_2-n_1} \Delta( \vec \Lambda) det( \{ e^{-\lambda_i w_1} \: (-\lambda_i)e^{-\lambda_i w_1} \: ... \: (-\lambda_i)^{n_1-2} e^{-\lambda_i w_1} \: e^{-\lambda_i w_2} \} )d \vec \Lambda.
\end{align*}
By following a procedure similar to the one followed for deriving capacity expressions, we can simplify the outage to,
\begin{align}\label{kmu_outage}
P(\lambda_{max}(W) <x)  &\approx (-1)^{\frac{1}{2} n_1(n_1-1)} \frac{1}{ \prod_{j=1}^{n_1} (n_2-j)! } \frac{1}{| \vec \Sigma|^{n_2} (w_2-w_1)^{n_1-1} \prod_{j=1}^{n_1-2} j!}det(\vec N).
\end{align}
where 
\begin{align*}
\vec N_{i,j}(n_1, n_2) =\begin{cases}
 \int_{0}^{x}  \lambda^{n_2-n_1} \lambda^{i-1} (-\lambda)^{j-1}  e^{-\lambda w_1} d\lambda;  \quad   1  \leq i \leq n_1 , 1 \leq j  \leq n_1-1\\
  \int_{0}^{x} \lambda^{n_2-n_1} \lambda^{i-1}e^{-\lambda w_2} d\lambda;  \quad   1  \leq i \leq n_1 , j= n_1.
 \end{cases}
\end{align*}
Using the identity $\int_{0}^{x} u^{v-1} e^{-\mu u} du= \mu^{-v} \gamma(v, \mu x)$, where $\gamma(.,.)$ is the incomplete gamma function, we obtain,
\begin{align*}
\vec N_{i,j}(n_1, n_2) =\begin{cases}
 (-1)^{j-1} w_1^{-n_2+n_1-i-j+1}\gamma(n_2-n_1+i+j-1, w_1x);  \quad   1  \leq i \leq n_1 , 1 \leq j  \leq n_1-1\\
w_2^{-n_2+n_1-i}\gamma(n_2-n_1+i, w_2x);  \quad   1  \leq i \leq n_1 , j= n_1.
 \end{cases}
\end{align*}
\bibliographystyle{IEEEtran}
\bibliography{bibfile}

\begin{thebibliography}{10}
\providecommand{\url}[1]{#1}
\csname url@samestyle\endcsname
\providecommand{\newblock}{\relax}
\providecommand{\bibinfo}[2]{#2}
\providecommand{\BIBentrySTDinterwordspacing}{\spaceskip=0pt\relax}
\providecommand{\BIBentryALTinterwordstretchfactor}{4}
\providecommand{\BIBentryALTinterwordspacing}{\spaceskip=\fontdimen2\font plus
\BIBentryALTinterwordstretchfactor\fontdimen3\font minus
  \fontdimen4\font\relax}
\providecommand{\BIBforeignlanguage}[2]{{%
\expandafter\ifx\csname l@#1\endcsname\relax
\typeout{** WARNING: IEEEtran.bst: No hyphenation pattern has been}%
\typeout{** loaded for the language `#1'. Using the pattern for}%
\typeout{** the default language instead.}%
\else
\language=\csname l@#1\endcsname
\fi
#2}}
\providecommand{\BIBdecl}{\relax}
\BIBdecl

\bibitem{james}
\BIBentryALTinterwordspacing
A.~T. James, ``Distributions of matrix variates and latent roots derived from
  normal samples,'' \emph{Ann. Math. Statist.}, vol.~35, no.~2, pp. 475--501,
  06 1964. [Online]. Available: \url{http://dx.doi.org/10.1214/aoms/1177703550}
\BIBentrySTDinterwordspacing

\bibitem{telatar_capacity}
\BIBentryALTinterwordspacing
E.~Telatar, ``{Capacity of Multi-antenna Gaussian Channels},'' \emph{European
  Transactions on Telecommunications}, vol.~10, no.~6, pp. 585--595, 1999.
  [Online]. Available: \url{http://dx.doi.org/10.1002/ett.4460100604}
\BIBentrySTDinterwordspacing

\bibitem{chiani_capacity}
M.~Chiani, M.~Win, and A.~Zanella, ``{On the capacity of spatially correlated
  MIMO Rayleigh-fading channels},'' \emph{IEEE Transactions on Information
  Theory}, vol.~49, no.~10, pp. 2363--2371, Oct 2003.

\bibitem{goldsmith_capacity}
A.~Goldsmith, S.~A. Jafar, N.~Jindal, and S.~Vishwanath, ``{Capacity limits of
  MIMO channels},'' \emph{IEEE Journal on Selected Areas in Communications},
  vol.~21, no.~5, pp. 684--702, June 2003.

\bibitem{kang_capacity}
M.~Kang and M.~S. Alouini, ``{Capacity of MIMO Rician channels},'' \emph{IEEE
  Transactions on Wireless Communications}, vol.~5, no.~1, pp. 112--122, Jan
  2006.

\bibitem{mckay_capacity}
M.~R. McKay and I.~B. Collings, ``{General Capacity Bounds for Spatially
  Correlated Rician MIMO Channels},'' \emph{IEEE Transactions on Information
  Theory}, vol.~51, no.~9, pp. 3121--3145, Sept 2005.

\bibitem{yacoub_k_mu}
M.~Yacoub, ``{The {\(\kappa\)}-{\(\mu\)} distribution and the
  {\(\eta\)}-{\(\mu\)} distribution},'' \emph{Antennas and Propagation
  Magazine, IEEE}, vol.~49, no.~1, pp. 68--81, Feb 2007.

\bibitem{capacity_costa}
D.~B.~D. Costa and M.~D. Yacoub, ``{Average channel capacity for generalized
  fading scenarios},'' \emph{IEEE Communications Letters}, vol.~11, no.~12, pp.
  949--951, December 2007.

\bibitem{outage_morales}
D.~Morales-Jimenez and J.~F. Paris, ``{Outage probability analysis for
  {\(\eta\)}-{\(\mu\)} fading channels},'' \emph{IEEE Communications Letters},
  vol.~14, no.~6, pp. 521--523, June 2010.

\bibitem{outage_paris}
J.~F. Paris, ``{Outage Probability in {\(\eta\)}-{\(\mu\)}/
  {\(\eta\)}-{\(\mu\)} and {\(\kappa\)}-{\(\mu\)} /{\(\eta\)}-{\(\mu\)}
  Interference-Limited Scenarios},'' \emph{IEEE Transactions on
  Communications}, vol.~61, no.~1, pp. 335--343, January 2013.

\bibitem{outage_ermolova}
N.~Y. Ermolova and O.~Tirkkonen, ``{Outage Probability Analysis in Generalized
  Fading Channels with Co-Channel Interference and Background Noise:
  {\(\eta\)}-{\(\mu\)}/ {\(\eta\)}-{\(\mu\)}, {\(\eta\)}-{\(\mu\)} /
  {\(\kappa\)}-{\(\mu\)} and {\(\kappa\)}-{\(\mu\)} / {\(\eta\)}-{\(\mu\)}
  Scenarios},'' \emph{IEEE Transactions on Wireless Communications}, vol.~13,
  no.~1, pp. 291--297, January 2014.

\bibitem{suman_coverage}
S.~Kumar and S.~Kalyani, ``{Coverage Probability and Rate for
  {\(\kappa\)}-{\(\mu\)} /{\(\eta\)}-{\(\mu\)} Fading Channels in
  Interference-Limited Scenarios},'' \emph{IEEE Transactions on Wireless
  Communications}, vol.~14, no.~11, pp. 6082--6096, Nov 2015.

\bibitem{suman_outage}
S.~Kumar, G.~Chandrasekaran, and S.~Kalyani, ``{Analysis of Outage Probability
  and Capacity for {\(\kappa\)}-{\(\mu\)} /{\(\eta\)}-{\(\mu\)} Faded
  Channel},'' \emph{IEEE Communications Letters}, vol.~19, no.~2, pp. 211--214,
  Feb 2015.

\bibitem{outage_mrc_morales}
D.~Morales-Jimenez, J.~F. Paris, and A.~Lozano, ``{Outage Probability Analysis
  for MRC in {\(\kappa\)}-{\(\mu\)} Fading Channels with Co-Channel
  Interference},'' \emph{IEEE Communications Letters}, vol.~16, no.~5, pp.
  674--677, May 2012.

\bibitem{srinivasan_optimum}
M.~{Srinivasan} and S.~{Kalyani}, ``Analysis of optimal combining in rician
  fading with co-channel interference,'' \emph{IEEE Transactions on Vehicular
  Technology}, vol.~68, no.~4, pp. 3613--3628, April 2019.

\bibitem{secrecy_bhargav}
N.~Bhargav, S.~L. Cotton, and D.~E. Simmons, ``{Secrecy Capacity Analysis Over
  {\(\kappa\)}-{\(\mu\)} Fading Channels: Theory and Applications},''
  \emph{IEEE Transactions on Communications}, vol.~64, no.~7, pp. 3011--3024,
  July 2016.

\bibitem{srinivasan_secrecy}
M.~{Srinivasan} and S.~{Kalyani}, ``Secrecy capacity of {\(\kappa\)}-{\(\mu\)}
  shadowed fading channels,'' \emph{IEEE Communications Letters}, vol.~22,
  no.~8, pp. 1728--1731, Aug 2018.

\bibitem{throughput_zhang}
J.~Zhang, Z.~Tan, H.~Wang, Q.~Huang, and L.~Hanzo, ``{The Effective Throughput
  of MISO Systems Over {\(\kappa\)}-{\(\mu\)} Fading Channels},'' \emph{IEEE
  Transactions on Vehicular Technology}, vol.~63, no.~2, pp. 943--947, Feb
  2014.

\bibitem{throughput_zhang1}
J.~Zhang, M.~Matthaiou, Z.~Tan, and H.~Wang, ``{Effective rate analysis of MISO
  {\(\eta\)}-{\(\mu\)} fading channels},'' in \emph{2013 IEEE International
  Conference on Communications (ICC)}, June 2013, pp. 5840--5844.

\bibitem{df_li}
W.~G. Li and M.~Chen, ``{Outage capacity of dual-hop decode-and-forward
  relaying system over generalized fading channels},'' in \emph{2010 2nd
  International Conference on Future Computer and Communication}, vol.~3, May
  2010, pp. V3--827--V3--831.

\bibitem{df_fikadu}
M.~K. Fikadu, P.~C. Sofotasios, S.~Muhaidat, Q.~Cui, and M.~Valkama, ``{SER of
  M-QAM decode-and-forward multi-relay systems under generalized fading
  conditions},'' in \emph{2016 23rd International Conference on
  Telecommunications (ICT)}, May 2016, pp. 1--5.

\bibitem{df_kumar}
P.~Kumar and K.~Dhaka, ``{Performance Analysis of a Decode-and-Forward Relay
  System in {\(\kappa\)}-{\(\mu\)} and {\(\eta\)}-{\(\mu\)} Fading Channels},''
  \emph{IEEE Transactions on Vehicular Technology}, vol.~65, no.~4, pp.
  2768--2775, April 2016.

\bibitem{athira_evt}
A.~{Subhash}, M.~{Srinivasan}, and S.~{Kalyani}, ``Asymptotic maximum order
  statistic for sir in {\(\kappa\)}-{\(\mu\)} shadowed fading,'' \emph{IEEE
  Transactions on Communications}, pp. 1--1, 2019.

\bibitem{kumar_random}
S.~Kumar and A.~Pandey, ``{Random Matrix Model for Nakagami-Hoyt Fading},''
  \emph{IEEE Transactions on Information Theory}, vol.~56, no.~5, pp.
  2360--2372, May 2010.

\bibitem{gholizadeh_capacity}
M.~H. Gholizadeh, H.~Amindavar, and J.~A. Ritcey, ``{On the capacity of MIMO
  correlated Nakagami-m fading channels using copula},'' \emph{EURASIP Journal
  on Wireless Communications and Networking}, vol. 2015, no.~1, p.~1, 2015.

\bibitem{mimo_vergara}
V.~M. Vergara and S.~E. Barbin, ``{MIMO capacity upper bound for
  {\(\kappa\)}-{\(\mu\)} and {\(\eta\)}-{\(\mu\)} faded channels},'' in
  \emph{2012 IEEE Radio and Wireless Symposium}, Jan 2012, pp. 367--370.

\bibitem{alfano_mimo}
G.~Alfano, A.~D. Maio, and A.~M. Tulino, ``{A Theoretical Framework for LMS
  MIMO Communication Systems Performance Analysis},'' \emph{IEEE Transactions
  on Information Theory}, vol.~56, no.~11, pp. 5614--5630, Nov 2010.

\bibitem{pozas_mimo}
L.~Moreno-Pozas and E.~Martos-Naya, ``{On Some Unifications Arising from the
  MIMO Rician Shadowed Model},'' in \emph{2016 IEEE 83rd Vehicular Technology
  Conference (VTC Spring)}, May 2016, pp. 1--5.

\bibitem{k_mu_phase}
U.~S. Dias, M.~D. Yacoub, and D.~B. da~Costa, ``{The {\(\kappa\)}-{\(\mu\)};
  phase-envelope joint distribution},'' in \emph{2008 IEEE 19th International
  Symposium on Personal, Indoor and Mobile Radio Communications}, Sept 2008,
  pp. 1--5.

\bibitem{eta_mu_phase}
D.~B. da~Costa and M.~D. Yacoub, ``{The {\(\eta\)}-{\(\mu\)}; Joint
  Phase-Envelope Distribution},'' in \emph{2007 IEEE Wireless Communications
  and Networking Conference}, March 2007, pp. 1906--1908.

\bibitem{kollo_wishart}
T.~Kollo and D.~von Rosen, ``{Approximating by the Wishart distribution},''
  \emph{Annals of the Institute of Statistical Mathematics}, vol.~47, no.~4,
  pp. 767--783, 1995.

\bibitem{tan_wishart}
W.~Tan and R.~Gupta, ``{On approximating the non-central wishart distribution
  by central wishart distribution a monte carlo study},'' \emph{Communications
  in Statistics-Simulation and Computation}, vol.~11, no.~1, pp. 47--64, 1982.

\bibitem{steyn}
H.~S. Steyn and J.~J.~J. Roux, ``{ Approximations for the non-central Wishart
  distribution},'' \emph{South African Statistical Journal}, vol.~6, no.~2, pp.
  165--172, Jan 1962.

\bibitem{pozas_shadowed}
L.~Moreno-Pozas, F.~J. Lopez-Martinez, J.~F. Paris, and E.~Martos-Naya, ``{The
  {\(\kappa\)}-{\(\mu\)} shadowed fading model: unifying the
  {\(\kappa\)}-{\(\mu\)} and {\(\eta\)}-{\(\mu\)} distributions},'' \emph{IEEE
  Trans. Veh. Tech.}, vol.~65, no.~12, pp. 9630--9641, Dec 2016.

\bibitem{cotton_d2d}
S.~L. Cotton, ``{Human body shadowing in cellular device-to-device
  communications: channel modeling using the shadowed {\(\kappa\)}-{\(\mu\)}
  fading model},'' \emph{IEEE Journal of Sel. Topics in Comm.}, vol.~33, no.~1,
  pp. 111--119, Jan 2015.

\bibitem{fraidenraich_alphakappa}
G.~{Fraidenraich} and M.~D. {Yacoub}, ``The {\(\alpha\)}-{\(\kappa\)}-{\(\mu\)}
  and {\(\alpha\)}-{\(\eta\)}-{\(\mu\)} fading distributions,'' in \emph{2006
  IEEE Ninth International Symposium on Spread Spectrum Techniques and
  Applications}, Aug 2006, pp. 16--20.

\bibitem{pablo_alphakappamu}
\BIBentryALTinterwordspacing
P.~Ramirez{-}Espinosa, J.~M.~M. Moualeu, D.~B. da~Costa, and F.~J.
  L{\'{o}}pez{-}Mart{\'{\i}}nez, ``The {\(\alpha\)}-{\(\kappa\)}-{\(\mu\)}
  shadowed fading distribution: Statistical characterization and
  applications,'' \emph{CoRR}, vol. abs/1904.05587, 2019. [Online]. Available:
  \url{http://arxiv.org/abs/1904.05587}
\BIBentrySTDinterwordspacing

\bibitem{fraidenraich_capacity}
G.~Fraidenraich, O.~Leveque, and J.~M. Cioffi, ``{On the MIMO Channel Capacity
  for the Nakagami-m Channel},'' \emph{IEEE Transactions on Information
  Theory}, vol.~54, no.~8, pp. 3752--3757, Aug 2008.

\bibitem{randombook}
\BIBentryALTinterwordspacing
R.~Couillet and M.~Debbah, \emph{Random Matrix Methods for Wireless
  Communications}.\hskip 1em plus 0.5em minus 0.4em\relax Cambridge University
  Press, 2011. [Online]. Available:
  \url{https://books.google.co.in/books?id=\_j7pT9HjKAUC}
\BIBentrySTDinterwordspacing

\bibitem{granstrom}
K.~Granstrom and U.~Orguner, ``{Properties and approximations of some matrix
  variate probability density functions},'' 2011.

\bibitem{boyd2004convex}
\BIBentryALTinterwordspacing
S.~Boyd, S.~Boyd, L.~Vandenberghe, and C.~U. Press, \emph{Convex Optimization},
  ser. Berichte {\"u}ber verteilte messysteme.\hskip 1em plus 0.5em minus
  0.4em\relax Cambridge University Press, 2004. [Online]. Available:
  \url{https://books.google.co.in/books?id=mYm0bLd3fcoC}
\BIBentrySTDinterwordspacing

\bibitem{kang2003mrc}
{Ming Kang} and M.~. {Alouini}, ``{Largest eigenvalue of complex Wishart
  matrices and performance analysis of MIMO MRC systems},'' \emph{IEEE Journal
  on Selected Areas in Communications}, vol.~21, no.~3, pp. 418--426, April
  2003.

\bibitem{marzetta2015massive}
T.~L. Marzetta, ``Massive mimo: an introduction,'' \emph{Bell Labs Technical
  Journal}, vol.~20, pp. 11--22, 2015.

\bibitem{bjornson2016massive}
E.~Bj{\"o}rnson, E.~G. Larsson, and T.~L. Marzetta, ``Massive mimo: Ten myths
  and one critical question,'' vol.~54, no.~2, pp. 114--123, 2016.

\bibitem{hanif2018antenna}
M.~Hanif, H.-C. Yang, G.~Boudreau, E.~Sich, and H.~Seyedmehdi, ``Antenna subset
  selection for massive mimo systems: A trace-based sequential approach for sum
  rate maximization,'' vol.~20, no.~2, pp. 144--155, 2018.

\bibitem{gao2016antenna}
Y.~Gao and T.~Kaiser, ``Antenna selection in massive mimo systems: Full-array
  selection or subarray selection?'' July 2016, pp. 1--5.

\bibitem{fan2015adc}
L.~Fan, S.~Jin, C.~K. Wen, and H.~Zhang, ``{Uplink Achievable Rate for Massive
  MIMO Systems With Low-Resolution ADC},'' vol.~19, no.~12, pp. 2186--2189, Dec
  2015.

\bibitem{singh2009adc}
J.~Singh, O.~Dabeer, and U.~Madhow, ``On the limits of communication with
  low-precision analog-to-digital conversion at the receiver,'' vol.~57,
  no.~12, pp. 3629--3639, December 2009.

\bibitem{bai2013adc}
Q.~Bai, A.~Mezghani, and J.~A. Nossek, ``On the optimization of adc resolution
  in multi-antenna systems,'' Aug 2013, pp. 1--5.

\bibitem{orhan2015adc}
O.~Orhan, E.~Erkip, and S.~Rangan, ``Low power analog-to-digital conversion in
  millimeter wave systems: Impact of resolution and bandwidth on performance,''
  Feb 2015, pp. 191--198.

\bibitem{zhang2016adc}
J.~Zhang, L.~Dai, S.~Sun, and Z.~Wang, ``{On the Spectral Efficiency of Massive
  MIMO Systems With Low-Resolution ADCs},'' vol.~20, no.~5, pp. 842--845, May
  2016.

\bibitem{tan2016adc}
W.~Tan, S.~Jin, C.~K. Wen, and Y.~Jing, ``{Spectral Efficiency of Mixed-ADC
  Receivers for Massive MIMO Systems},'' \emph{IEEE Access}, vol.~4, pp.
  7841--7846, 2016.

\bibitem{zhang2017adc}
J.~Zhang, L.~Dai, Z.~He, S.~Jin, and X.~Li, ``{Performance Analysis of
  Mixed-ADC Massive MIMO Systems Over Rician Fading Channels},'' vol.~35,
  no.~6, pp. 1327--1338, June 2017.

\bibitem{srinivasan2019analysis}
M.~Srinivasan and S.~Kalyani, ``Analysis of massive mimo with low resolution
  adc in nakagami-m fading,'' \emph{IEEE Commun. Lett.}, 2019.

\bibitem{ding2018mimo}
Q.~Ding and Y.~Jing, ``{Outage Probability Analysis and Resolution Profile
  Design for Massive MIMO Uplink With Mixed-ADC},'' vol.~17, no.~9, pp.
  6293--6306, Sept 2018.

\bibitem{qiao_adc}
D.~{Qiao}, W.~{Tan}, Y.~{Zhao}, C.~{Wen}, and S.~{Jin}, ``Spectral efficiency
  for massive mimo zero-forcing receiver with low-resolution adc,'' in
  \emph{2016 8th International Conference on Wireless Communications Signal
  Processing (WCSP)}, Oct 2016, pp. 1--6.

\bibitem{jacobsson_adc}
S.~{Jacobsson}, G.~{Durisi}, M.~{Coldrey}, U.~{Gustavsson}, and C.~{Studer},
  ``Throughput analysis of massive mimo uplink with low-resolution adcs,''
  \emph{IEEE Transactions on Wireless Communications}, vol.~16, no.~6, pp.
  4038--4051, June 2017.

\bibitem{liuarxiv}
T.~{Liu}, J.~{Tong}, Q.~{Guo}, J.~{Xi}, Y.~{Yu}, and Z.~{Xiao}, ``{On the
  Performance of Massive MIMO Systems With Low-Resolution ADCs Over Rician
  Fading Channels},'' \emph{arXiv e-prints}, Jun 2019.

\bibitem{reyes}
M.~A. Reyes and R.~Arcos-Olalla, ``{Supersymmetric features of the Error and
  Dawson's functions},'' \emph{Revista mexicana de f{\'\i}sica}, vol.~61,
  no.~6, pp. 475--480, 2015.

\bibitem{abr}
M.~Abramowitz and I.~Stegun, \emph{Handbook of Mathematical Functions with
  Formulas, Graphs, and Mathematical Tables}, ser. Applied mathematics
  series.\hskip 1em plus 0.5em minus 0.4em\relax U.S. Government Printing
  Office, 1972.

\bibitem{ng_table}
E.~W. Ng and M.~Geller, ``{A table of integrals of the error functions},''
  \emph{Journal of Research of the National Bureau of Standards B}, vol.~73,
  no.~1, pp. 1--20, 1969.

\bibitem{for}
\BIBentryALTinterwordspacing
I.~Wolfram~Research. Generalized laguerre polynomials. [Online]. Available:
  \url{http://functions.wolfram.com/07.23.16.0002.01}
\BIBentrySTDinterwordspacing

\bibitem{brychkov}
Y.~A. Brychkov and N.~Saad, ``{On some formulas for the Appell function F2 (a,
  b, b'; c, c'; w; z)},'' \emph{Integral Transforms and Special Functions},
  vol.~25, no.~2, pp. 111--123, 2014.

\bibitem{for2}
\BIBentryALTinterwordspacing
I.~Wolfram~Research. Generalized laguerre polynomials. [Online]. Available:
  \url{http://functions.wolfram.com/01.04.26.0003.01}
\BIBentrySTDinterwordspacing

\bibitem{int}
A.~Jeffrey and D.~Zwillinger, \emph{Table of Integrals, Series, and Products},
  ser. Table of Integrals, Series, and Products Series.\hskip 1em plus 0.5em
  minus 0.4em\relax Elsevier Science, 2007.

\end{thebibliography}
\end{document}